\documentclass[11pt]{article}

\usepackage[margin=0.93in]{geometry}
\usepackage{setspace}
\usepackage{parskip}
\usepackage{pdflscape}
\usepackage{etex}

\usepackage[utf8]{inputenc}
\usepackage{times}
\usepackage{authblk}

\usepackage{amsfonts,latexsym,amsthm,amssymb,amsmath,amscd,euscript,bbm,bm}
\usepackage{mathtools}
\usepackage{nicefrac, xfrac}

\usepackage[authoryear]{natbib}
\renewcommand{\cite}{\citep}
\usepackage[resetlabels]{multibib}
\newcites{SI}{References}
\usepackage{sidecap}
\sidecaptionvpos{figure}{c}
\usepackage[bf,margin=0.5cm,font=footnotesize]{caption} 

\usepackage{multirow}
\usepackage{longtable}
\usepackage{booktabs}
\usepackage{colortbl}
\usepackage{blkarray}
\usepackage{tabularx}

\usepackage[dvipsnames]{xcolor}
\usepackage[usenames,dvipsnames]{pstricks}

\usepackage[hypertexnames=false]{hyperref}
\hypersetup{
    pdftitle = {Kawakatsu, Kessinger, Plotkin\_2023\_A mechanistic model of gossip, reputations, and cooperation},
    pdfauthor = {Mari Kawakatsu, Taylor Kessinger, Joshua Plotkin}
    hidelinks,
    colorlinks = true,
    bookmarksnumbered,
    citecolor = black,
    linkcolor = black,
    urlcolor  = black
    }
    
\usepackage[capitalise,nameinlink]{cleveref}
\usepackage{autonum}
\creflabelformat{equation}{#2#1#3}
\usepackage{nameref}

\usepackage{pifont}
\usepackage{comment}
\usepackage{lastpage}
\usepackage{fancyhdr}
\usepackage{lineno}

\usepackage{titlesec}
\newcommand{\addperiod}[1]{#1.}
\titleformat{\paragraph}[runin]{\normalsize\bfseries}{\thesection}{1em}{\addperiod}
\titleformat{\section}{\Large\bfseries}{\thesection}{0.5em}{}
\titleformat{\subsection}{\large\bfseries}{\thesubsection}{0.5em}{}
\titleformat{\subsubsection}{\normalsize\bfseries}{\thesubsubsection}{0.5em}{}
\titlespacing*{\section}{0pt}{3ex plus .5ex minus .2ex}{1.5ex plus .1ex}
\titlespacing*{\subsection}{0pt}{2.5ex plus .5ex minus .2ex}{1.5ex plus .1ex}
\titlespacing*{\subsubsection}{0pt}{2.5ex plus .5ex minus .2ex}{1.5ex plus .1ex}
\titlespacing*{\paragraph}{0pt}{\parskip}{1\parskip}

\newcommand{\strat}[0]{s}
\newcommand{\allc}[0]{{\textrm{ALLC}}}
\newcommand{\alld}[0]{{\textrm{ALLD}}}
\newcommand{\disc}[0]{{\textrm{DISC}}}

\newcommand{\exec}{u_e}
\newcommand{\assess}{u_a}
\newcommand{\p}{p}
\newcommand{\q}{q}
\newcommand{\pgc}[0]{P_{GC}}
\newcommand{\pgd}[0]{P_{GD}}
\newcommand{\pbc}[0]{P_{BC}}
\newcommand{\pbd}[0]{P_{BD}}

\newcommand{\GtoB}[0]{u}
\newcommand{\BtoG}[0]{v}
\newcommand{\GTOB}[0]{\mu}
\newcommand{\BTOG}[0]{\nu}

\newcommand{\queen}[0]{Q}
\newcommand{\sgen}[0]{\tau}
\newcommand{\gen}[0]{T}
\newcommand{\rtil}[0]{\tilde r}
\newcommand{\gtwo}[0]{\tilde g_2}
\newcommand{\btwo}[0]{\tilde b_2}
\newcommand{\dtwo}[0]{\tilde d_2}

\DeclareMathOperator{\Var}{Var}
\DeclareMathOperator{\E}{\mathbb{E}}

\newcommand{\SI}[0]{{\color{black}{Supplementary Information}}}
\newcommand{\maintext}[0]{{\color{black}{main text}}}

\title{\vspace{1ex}A mechanistic model of gossip, reputations, and cooperation}
\date{\today}
\author[1,2,*,$\dagger$]{Mari Kawakatsu}
\author[1,*,$\dagger$]{Taylor A. Kessinger}
\author[1,2,$\dagger$]{Joshua B. Plotkin}
\affil[1]{\small Department of Biology, University of Pennsylvania, Philadelphia, PA 19104}
\affil[2]{\small Center for Mathematical Biology, University of Pennsylvania, Philadelphia, PA 19104}
\affil[*]{\small These authors contributed equally}
\affil[$\dagger$]{\small Corresponding authors: \href{mailto:marikawa@sas.upenn.edu}{marikawa@sas.upenn.edu} (M.K.), \href{mailto:tkess@sas.upenn.edu}{tkess@sas.upenn.edu} (T.A.K.), \href{mailto:jplotkin@sas.upenn.edu}{jplotkin@sas.upenn.edu}~(J.B.P.)}

\begin{document}
\maketitle


\pagenumbering{arabic}
\fancyhf{}
\fancypagestyle{plain}{
    \setcounter{page}{1}
    \fancyhf{}
    \fancyfoot[R]{\small Page \thepage \hspace{1pt} of \pageref{end}}
    \renewcommand{\headrulewidth}{0pt}
    \renewcommand{\footrulewidth}{0pt}}
\pagestyle{plain}


\begin{abstract}

    \noindent Social reputations facilitate cooperation: those who help others gain a good reputation, making them more likely to receive help themselves. But when people hold private views of one another, this cycle of indirect reciprocity breaks down, as disagreements lead to the perception of unjustified behavior that ultimately undermines cooperation. Theoretical studies often assume population-wide agreement about reputations, invoking rapid gossip as an endogenous mechanism for reaching consensus. However, the theory of indirect reciprocity lacks a mechanistic description of how gossip actually generates consensus. Here we develop a mechanistic model of gossip-based indirect reciprocity that incorporates two alternative forms of gossip: exchanging information with randomly selected peers or consulting a single gossip source. We show that these two forms of gossip are mathematically equivalent under an appropriate transformation of parameters. We derive an analytical expression for the minimum amount of gossip required to reach sufficient consensus and stabilize cooperation. We analyze how the amount of gossip necessary for cooperation depends on the benefits and costs of cooperation, the assessment rule (social norm), and errors in reputation assessment, strategy execution, and gossip transmission. Finally, we show that biased gossip can either facilitate or hinder cooperation, depending on the direction and magnitude of the bias. Our results contribute to the growing literature on cooperation facilitated by communication, and they highlight the need to study strategic interactions coupled with the spread of social information.
    
\end{abstract}


\vspace{10ex}
\section{Introduction}

Reputations and social norms are critical for cooperation in large human societies \cite{trivers_evolution_1971,alexander_biology_1987,tomasello_origins_2013}. Individuals can improve their reputations by behaving altruistically, making others more likely to help them in the future. According to a large body of theoretical work \cite{boyd_evolution_1989,nowak_dynamics_1998,leimar_evolution_2001,nowak_evolution_2005}, this feedback loop, termed indirect reciprocity, can maintain cooperation even among strangers. There is also ample empirical evidence that reputations facilitate altruistic behavior: in laboratory settings, people are more likely to offer help when others are observing them \cite{bereczkei_public_2007} or when others have knowledge of their behavioral history \cite{milinski_reputation_2002}; field studies show that individuals of higher social status are more likely to gain cooperative partners \cite{bliegebird_prosocial_2015,vonrueden_dynamics_2019}.

Indirect reciprocity facilitates cooperation only when individuals agree about each other's social standing. The standard theory of indirect reciprocity assumes by fiat that reputations are common knowledge so that the entire population agrees about the reputation of each individual \cite{nowak_evolution_1998,ohtsuki_how_2004,ohtsuki_leading_2006}. Consensus about reputations helps maintain cooperation, as individuals choose to cooperate with those of good social standing, thereby earning good reputations for themselves. 

However, when people hold private opinions about each other's social standing, disagreements can lead to the perception of unjustified behavior that eventually undermines cooperation \cite{uchida_effect_2010,uchida_effect_2013,okada_tolerant_2017,okada_solution_2018,hilbe_indirect_2018,ohtsuki_indirect_2009}. Theoretical studies have proposed several mechanisms that could help maintain cooperation even when reputations are held privately---including empathetic perspective taking \cite{radzvilavicius_evolution_2019}, generous moral evaluation \cite{schmid_evolution_2021}, nuanced quantitative assessments \cite{schmid_quantitative_2023}, or a monitoring system that broadcasts public information about reputations \cite{radzvilavicius_adherence_2021}. 

Nonetheless, the most common justification for assuming consensus about reputations \cite{nowak_evolution_2005,ohtsuki_indirect_2009,santos_social_2018,okada_tolerant_2017,okada_solution_2018,radzvilavicius_evolution_2019,radzvilavicius_adherence_2021,harrison_strength_2011,kessinger_evolution_2023,morsky_indirect_2023} is an endogenous mechanism of rapid gossip within a population---that is, the exchange of information about the social standings of others \cite{foster_research_2004,sommerfeld_gossip_2007,balliet_indirect_2020}. According to this reasoning, even if individuals initially disagree about each other's standing, rapid gossip will eventually lead to consensus. The role of gossip in cooperation also has empirical support, as laboratory \cite{wu_when_2015,beersma_how_2011,wu_reputation_2016b,sommerfeld_gossip_2007,wu_when_2015,feinberg_gossip_2014,wu_reputation_2016a,feinberg_virtues_2012} and field \cite{dorescruz_gossip_2021} studies show that people tend to cooperate more when (they believe) their peers gossip about their behavior. 

Despite the intuitive appeal of gossip and empirical studies of its effects, the theory of indirect reciprocity lacks a mechanistic description of how gossip produces consensus about social standings in a population. Existing work on gossip has focused on how gossip allows recipients to detect potential cheaters and selectively avoid them \cite[partner choice;][]{wu_when_2015,feinberg_gossip_2014,traag_indirect_2011,traag_dynamical_2013} or how honest or dishonest gossip can incentivize cooperation or punish free riders \cite[gossip strategies;][]{wu_reputation_2016a,feinberg_virtues_2012,nakamaru_evolution_2004,seki_model_2016,wu_honesty_2021}. But how gossip produces consensus about reputations---and thereby stabilizes cooperation---has received less attention \cite{ohtsuki_indirect_2009,righi_gossip_2022}. Several key questions remain unanswered: How much gossip is required to support cooperation? How does the structure of gossip transmission govern convergence to consensus? How will noise or bias in transmission deteriorate the effects of gossip?

Here we address these questions by developing a model of indirect reciprocity that integrates a mechanistic description of gossip about social reputations. We consider two forms of gossip: exchanging information with randomly selected peers or consulting a single gossip source. We show that these two gossip processes are mathematically equivalent under an appropriate transformation of parameters. We then derive an analytical expression for the minimum amount of gossip required to stabilize cooperation, and we discuss how this critical gossip duration depends on model parameters, including the benefit-to-cost ratio for cooperation, the assessment rule (social norm), and the rates of error in reputation assessment, strategy execution, and gossip transmission. We conclude by showing that biased gossip---that is, sharing false information about another individual's social standing---can either facilitate or hinder cooperation, depending on its direction (positive or negative) and magnitude.

\section{A model of gossip, reputations, and social behavior}
\label{sec:model}
\phantomsection

\subsection{Social Interactions} 

We build on a well-established framework for modeling cooperation by indirect reciprocity \cite{sasaki_evolution_2017,okada_solution_2018}. A large, well-mixed population of individuals engage in pairwise social interactions. Each interaction takes the form of a one-shot donation game. In each game, the \textit{donor} chooses whether or not to cooperate with the \textit{recipient} by paying a cost $c>0$ to provide a benefit $b>c$. If the donor defects, she incurs no cost and provides no benefit to the recipient.

Whether or not a donor cooperates depends on her current behavioral strategy. We consider the three strategies that are most common in studies of indirect reciprocity \cite{sasaki_evolution_2017,santos_social_2018,radzvilavicius_evolution_2019}: always cooperate ($\allc$), which means the donor intends to cooperate with any recipient; always defect ($\alld$), which means the donor defects against any recipient; and discriminate ($\disc$), which means the donor intends to cooperate when the recipient has a good reputation but defect when the recipient has a bad reputation. We allow for errors in strategy execution \cite{sasaki_evolution_2017,okada_solution_2018,hilbe_indirect_2018}: with probability $0<\exec<1/2$ (\textit{execution error rate}), a donor erroneously defects while intending to cooperate.

The resulting payoffs of cooperators ($\allc$), defectors ($\alld$), and discriminators ($\disc$) are given by
\begin{equation}
    \label{eq:payoffs}
    \begin{split}
        \pi_{\allc} & = \left(1-\exec\right) \Big[ b \left( f_{\allc} + f_{\disc} \cdot r_{\allc} \right)  - c \Big] \;,\\
        \pi_{\alld} & = \left(1-\exec\right) \Big[ b \left( f_{\allc} + f_{\disc} \cdot r_{\alld} \right) \Big] \;,\\
        \pi_{\disc} & = \left(1-\exec\right) \Big[ b \left( f_{\allc} + f_{\disc} \cdot r_{\disc} \right) - c \cdot r  \Big] \;,
    \end{split}
\end{equation}
where $f_{\strat}$ is the frequency of strategic type $\strat\in S=\{\allc,\alld,\disc\}$ in the population, satisfying $\sum_{\strat\in S} f_\strat = 1$. Here $r_\strat$ denotes the \textit{average reputation of type $s$}, i.e.~the fraction of the population that views an individual of type $\strat$ as good; and $r = \sum_{\strat\in S}f_{\strat} \cdot r_{\strat}$ is the \textit{average reputation} in the population.

\begin{figure}[b!]
    \centering
    \includegraphics[width=0.92\linewidth]{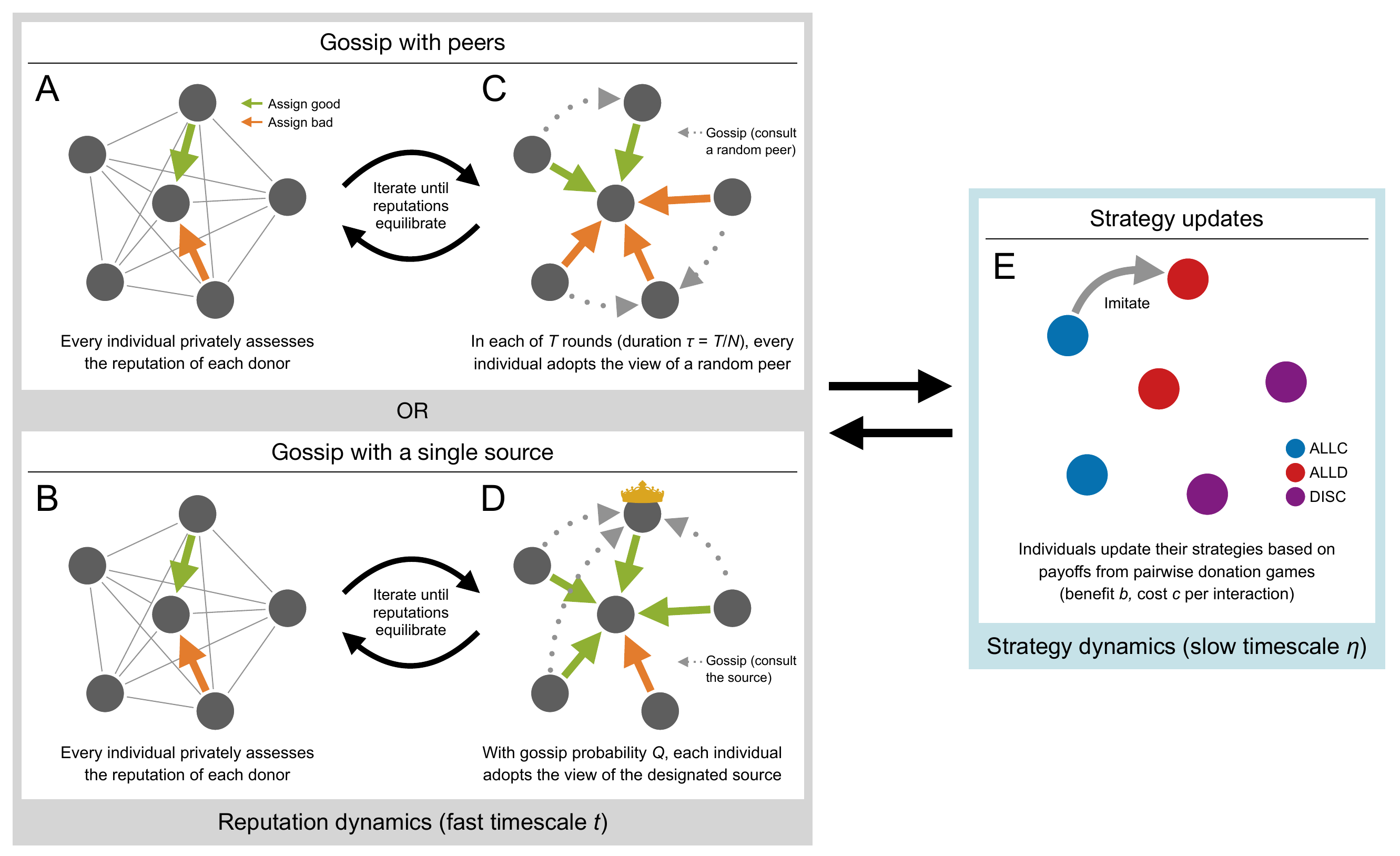} 
    \caption{
        \textbf{A model of gossip, reputations, and social behavior.}
        We consider a large, well-mixed population of individuals (nodes) engaged in pairwise social interactions (edges).
        \textbf{A, B}: After a round of pairwise social interactions, individuals privately assess each donor's reputation by judging her action toward a randomly selected recipient. 
        As a result of independent observations, individuals may disagree about the reputation of a given donor (orange and green arrows).
        \textbf{C, D}: Private assessments are followed by a period of gossip, governed by one of two mechanisms.
        \textbf{C}: \textit{Gossip with peers}. In each of $\gen$ rounds (equivalent to scaled duration $\sgen=\gen/N$, where $N$ is the population size), each individual consults a randomly selected peer (dotted arrows) and adopts her view of the focal individual.
        \textbf{D}: \textit{Gossip with a single source}. With probability $\queen$, each observer consults the same, designated gossip source (dotted arrows towards the node with the yellow crown) and adopts her view of the focal individual.
        \textbf{E}: Private observations and periods of gossip (steps A and C or steps B and D) repeat until reputations equilibrate. Once reputations reach equilibrium, individuals update their behavioral strategies by payoff-biased imitation. Colors indicate three possible behavioral strategies: $\allc$ (blue), $\alld$ (red), and $\disc$ (purple).
    }
    \label{fig:schematic}
\end{figure}

\subsection{Reputation updates (fast timescale)}
After a round of pairwise game play---that is, after every individual interacts with every other individual, serving once as a donor and once as a recipient---individuals then privately assess the reputation of each donor by observing her action toward a randomly selected recipient (\cref{fig:schematic}A, B). At this point, individuals may disagree about the reputation of a given donor because they assessed the donor based on her interaction with potentially different recipients. In addition, we assume there is a small probability of error $0<\assess<1/2$ (\textit{assessment error rate}) for each assessment \cite{ohtsuki_global_2007,hilbe_indirect_2018}, which occurs independently for each person who assesses a donor. 

Private assessments are then followed by a period of gossip about reputations (\cref{fig:schematic}C or D), which tends to increase agreement (see below). After the gossip period, there is yet another round of private assessments. Subsequent periods of private assessments and periods of gossip occur iteratively until reputations equilibrate. 

Under these assumptions, the average reputation of each strategic type $\strat$ changes according to the ODEs \cite[see \nameref{sec:methods};][]{perret_evolution_2021},
\begin{equation}
    \dfrac{dr_\strat}{dt} = p_\strat(t) - r_\strat(t), \qquad \strat\in S=\{\allc,\alld,\disc\} \;,
    \label{eq:repODE}
\end{equation}
where $p_\strat(t)$ is the probability that an individual of strategic type $\strat$ will be assigned a good reputation by an observer, which depends on the current reputations in the population as follows (see \nameref{sec:methods}):
\begin{equation}
    \label{eq:rep_ps}
    \begin{split}
        p_{\allc}(t) & = r(t) \pgc + \left(1-r(t)\right) \pbc \;, \\
        p_{\alld}(t) & = r(t) \pgd + \left(1-r(t)\right) \pbd \;, \\
        p_{\disc}(t) & = \gtwo(t) \pgc + \dtwo(t) \left(\pbc + \pgd\right) + \btwo(t) \pbd \;,\\
        r(t)  & \triangleq \sum_{\strat\in S}f_{\strat} \cdot r_{\strat}(t)
        = f_{\allc} \cdot r_{\allc}(t) + f_{\alld} \cdot r_{\alld}(t) + f_{\disc} \cdot r_{\disc}(t) \;.
    \end{split}
\end{equation}
Here, the expression $\gtwo$ denotes the probability that, after a period of gossip, two randomly selected individuals agree that a third individual is good; $\btwo$ denotes the probability that, after gossip, two individuals agree that a third individual is bad; and $2\dtwo$ denotes the probability that, after gossip, two individuals disagree about the reputation of a third. This dynamical system for the average reputation of the three strategic types has the same form as in prior models of private reputations \cite{radzvilavicius_evolution_2019,radzvilavicius_adherence_2021,kessinger_evolution_2023}, except that the terms $\gtwo$, $\dtwo$, and $\btwo$ differ from prior studies as the result of the gossip process, as we will describe below. 

We refer to $r_\strat$ evaluated at the equilibrium of \cref{eq:repODE} as the \textit{equilibrium reputation} of strategic type $\strat$, and we compute the \textit{agreement level} at the reputation equilibrium as $\gtwo+\btwo=\sum_\strat f_\strat r_\strat^2 + \sum_\strat f_\strat \left(1-r_\strat\right)^2$. As we will see below, gossip does not change the reputation dynamics or the equilibrium reputations for individuals using strategy $\alld$ or $\allc$, but it will tend to increase the level of agreement in the population about such reputations. By contrast, gossip can change the equilibrium reputation of discriminators ($r_\disc$) in the population, in addition to increasing the level of agreement about such reputations. 

The quantities $P_{XY}$ in \cref{eq:rep_ps} correspond to the assessment stage in the iterated rounds of private assessments and gossip. In particular, $P_{XY}$ denotes the probability that an observer will assign a focal individual a good reputation after she takes action $X\in\{\textrm{Cooperate }(C), \textrm{Defect }(D)\}$ against a recipient who has reputation $Y\in\{\textrm{Good }(G), \textrm{Bad }(B)\}$ in the eyes of the observer. This quantity is dictated by the \textit{social norm}, a set of assessment rules that govern how an observer judges a donor's reputation (good or bad) based on her action (cooperate or defect) toward a recipient \cite{ohtsuki_how_2004,ohtsuki_leading_2006,nowak_evolution_2005,hilbe_indirect_2018,santos_social_2018}. We consider three second-order social norms that are most common in studies of indirect reciprocity \cite{sasaki_evolution_2017,radzvilavicius_evolution_2019,radzvilavicius_adherence_2021,kessinger_evolution_2023}: Stern Judging, Simple Standing, and Shunning (see \nameref{sec:methods}).

\subsection{Gossip} 

The dynamics described above include terms that account for gossip, which tends to increase agreement about reputations. We develop models for two forms of gossip: pairwise gossip between random peers or gossip with a single source. We analyze how gossip modifies the level of agreement in the population about each other's reputations. 

In the absence of gossip, there are classical expressions for the probability that two independent private observers will agree a given focal individual is good ($g_2=\sum_{s\in S} f_\strat r_\strat^2$), agree a focal individual is bad ($b_2=\sum_{s\in S} f_\strat \left(1-r_\strat\right)^2$), or disagree about the reputation of a focal individual ($2d_2 =2\sum_{s\in S} f_\strat r_\strat \left(1-r_\strat\right)$). These expressions assume independent observations of the focal individual's action towards a random recipient \cite{radzvilavicius_evolution_2019,kessinger_evolution_2023}. By contrast, the levels of agreement and disagreement after a period of gossip, denoted $\gtwo$, $\btwo$, and $\dtwo$, will depart from the classical case of independent assessment, as described below.

\textbf{Peer-to-peer gossip.}
We model gossip as a process in which the reputation of a focal individual (the subject of gossip) spreads from peer to peer (\cref{fig:schematic}C). We consider a finite population of $N$ individuals engaged in gossip. At each round of $\gen$ rounds during this gossip process, and for each focal individual $i$, every individual randomly selects a peer and adopts her view of $i$'s reputation. The gossip dynamics for a focal individual are therefore described by a bi-allelic haploid Wright-Fisher process, which keeps track of how many individuals view the focal individual as good (allele one) or bad (allele two) over discrete generations (rounds) of gossip. In each generation, the choice of the peer from whom to receive gossip corresponds to the choice of parentage in a neutral coalescent (see \nameref{sec:methods}). The Wright-Fisher processes describing gossip about different focal individuals are assumed independent.

At the start of the gossip process, the fraction $r_i$ of the population who view a focal individual $i$ of type $\strat$ as good is given by fraction $r_\strat(t)$ of the population who view type $\strat$ as good in the context of the reputation ODEs that track the average reputations of different types (\cref{eq:repODE}). After $\gen$ rounds (Wright-Fisher generations) of peer-to-peer gossip, the agreement and disagreement terms are modified as follows:
\begin{equation}
    \label{eq:g2b2d2-wf-largeN} 
    \begin{split}  
        \gtwo & = g_2 + d_2 \cdot \left(1-e^{-\sgen} \right) \;,\\
        \btwo & = b_2 + d_2 \cdot \left(1-e^{-\sgen} \right) \;,\\
        \dtwo & = d_2 \cdot e^{-\sgen} \;,
    \end{split}
\end{equation}
where we define $\sgen\triangleq\gen/N$ as the scaled gossip duration, and where again $g_2$, $b_2$, $d_2$ denote corresponding agreement and disagreement terms from private observations before gossip \cite{radzvilavicius_evolution_2019,kessinger_evolution_2023}. These expressions for the effect of gossip are derived from the loss of heterozygosity over time in a Wright-Fisher process (\nameref{sec:methods}). 

The number of gossip rounds, $\gen$, quantifies the amount of peer-to-peer gossip that occurs in between periods of private observations. Thus, the duration of each gossip period, $\gen$, can be thought of as the relative rate of gossip versus private information. The case $\sgen \rightarrow \infty$ (infinitely long period of peer-to-peer gossip) is equivalent to public information about reputations, with no disagreements ($\dtwo=0$).

\textbf{Gossip with a single source.}
As an alternative model of gossip, we consider information transfer with a single source. In this case, we suppose that in each period of gossip, a randomly selected individual serves as the sole source of gossip (\cref{fig:schematic}D). Each individual then decides either to retain their private view of a donor's reputation (with probability $1-\queen$, $0\leq Q\leq 1$) or to consult the gossip source (with probability $\queen$) and adopt the source's view of the donor. Decisions on whether or not to consult the source are made independently for each individual's view of each individual. 

The resulting rates of agreement and disagreement after a period of single-source gossip are given by (\nameref{sec:methods})
\begin{equation}
    \label{eq:g2b2d2-queen}
    \begin{split}
        \gtwo & = \left(1-\queen^2\right) \cdot g_2 + \queen^2 \cdot r \;, \\
        \btwo & = \left(1-\queen^2\right) \cdot b_2 + \queen^2 \cdot (1-r) \;, \\
        \dtwo & = \left(1-\queen^2\right) \cdot d_2 \;.
    \end{split}
\end{equation}
Here, $\queen^2$ represents the probability that a random observer and a random donor both consulted the gossip source. The quantity $\queen^2$ is mathematically equivalent to the probability of unilateral empathetic assessment studied in \citet{radzvilavicius_evolution_2019} (\SI). 

Note that the case $Q =1$ (assured consultation of the gossip source) is equivalent to public information about reputations, with no disagreements. (In fact, since $r=g_2+d_2$ and $1-r=b_2+d_2$, the expressions for $\gtwo$, $\btwo$, and $\dtwo$ in \cref{eq:g2b2d2-queen} match the corresponding expressions in \cref{eq:g2b2d2-wf-largeN} in the limit of public information, i.e.~$\queen=1$ or $\sgen\rightarrow\infty$.)

\subsection{Strategy updates (slow timescale)}
Reputations change through iterated rounds of private observations and periods of gossip, eventually reaching equilibrium values for each strategic type, given by the equilibrium of \cref{eq:repODE}. After reputations equilibrate, individuals then update their strategies by payoff-biased imitation \cite[\cref{fig:schematic}E;][]{hofbauer_evolutionary_1998}. This modeling framework assumes a separation of timescales, motivated by the idea that reputations change quickly, whereas people are slow to change their behavior. That is, we assume that reputations equilibrate before individuals update behavioral strategies, as is standard in studies of indirect reciprocity \cite{uchida_effect_2010,okada_solution_2018,sasaki_evolution_2017,hilbe_indirect_2018}.

We describe the dynamics of competing strategies using replicator-dynamic ODEs \cite{taylor_evolutionary_1978},
\begin{equation}
    \dfrac{df_\strat}{d\eta} = f_\strat(\eta) \left(\pi_\strat(\eta) - \bar \pi\right(\eta)) \;,
    \label{eq:replicator}
\end{equation}
where $\pi_\strat(\eta)$ denotes the payoff to an individual of strategic type $s$ (\cref{eq:payoffs}) and $\bar \pi (\eta)= \sum_{\strat\in S} f_\strat(\eta) \pi_\strat(\eta)$ denotes the average payoff of the population, at time $\eta$. We use a different notation for time, $\eta$, to describe the strategy dynamics in order to distinguish this process from the reputation dynamics. The reputation dynamics occur on a faster timescale, denoted $t$, and they reach equilibrium (and influence payoffs) before any strategic changes occur.

\section{Results}
\phantomsection
\subsection{Gossip with a single source is equivalent to peer-to-peer gossip} 

Both proposed mechanisms of gossip---consulting a single source or transferring reputation information between peers---will tend to increase agreement about reputations across the population. To gain some intuition for this effect, we will start by comparing the two models of gossip to one another before considering their downstream impact on behavioral evolution.

The duration of peer-to-peer gossip ($\sgen$) governs the extent of agreement that peer-to-peer gossip induces, as does the probability of consulting the source ($\queen$) under the single-source gossip model. By comparing the expressions for $\gtwo$, $\dtwo$, and $\btwo$ in \cref{eq:g2b2d2-wf-largeN} (peer-to-peer model; \cref{fig:schematic}A, C) and \cref{eq:g2b2d2-queen} (single-source model; \cref{fig:schematic}B, D), we see that the two models of gossip are, in fact,  mathematically equivalent, with the following mapping between the duration of peer-to-peer gossip $\sgen$ and the probability $\queen$ of consulting the single source:
\begin{equation}
    \sgen = -\log(1-\queen^2) \;.
    \label{eq:mapping}
\end{equation}
The classical case of fully private information \cite{okada_solution_2018} corresponds to no peer-to-peer gossip ($\gen=0$) or, equivalently, to no consultation with the single source ($\queen=0$). By contrast, the case of fully public information \cite{okada_solution_2018} corresponds to the limit of an infinitely long duration of peer-to-peer gossip ($\gen\rightarrow\infty$) or, equivalently, assured consultation of the single source ($\queen\rightarrow1$); in this limit, there will be no disagreement about reputations ($\dtwo = 0$). Thus, these mechanistic models of gossip span continuously between public and private information about reputations.

\Cref{eq:mapping} provides some quantitative intuition about the relationship between single-source and peer-to-peer gossip (see also \cref{fig:mapping}). For example, peer-to-peer gossip for duration $\sgen=1/100$ (e.g., one Wright-Fisher generation in a population of $100$ individuals, or $100$ peer-to-peer gossip events) corresponds to single-source gossip with $\queen = 0.0998$ (i.e., $\approx$ 10\% chance of consulting the gossip source \textit{per individual}). Whereas peer-to-peer gossip for duration $\sgen=1$ ($N$ Wright-Fisher generations in a population of $N$ individuals, i.e., $N^2$ peer-to-peer gossip events) corresponds to $\queen = 0.795$ (i.e., $\approx$ 80\% chance of consulting the single gossip source \textit{per individual}).

\subsection{Gossip stabilizes cooperation}

We will use the peer-to-peer model (\cref{eq:g2b2d2-wf-largeN}) to study how gossip impacts reputations and cooperation. All our results can be translated into the language of the single-source model using the transformation given by \cref{eq:mapping}. We focus on understanding how gossip can stabilize cooperation under the Stern Judging norm---because this norm provides the highest rates of cooperation under public information, but it renders cooperation vulnerable to invasion by defectors when reputations are assessed privately without any gossip \cite{okada_solution_2018}. In \SI, we report corresponding results for the Simple Standing and Shunning norms (\cref{fig:stabilityALL}).

\begin{figure}[h!]
    \centering
    \includegraphics[width=0.92\linewidth]{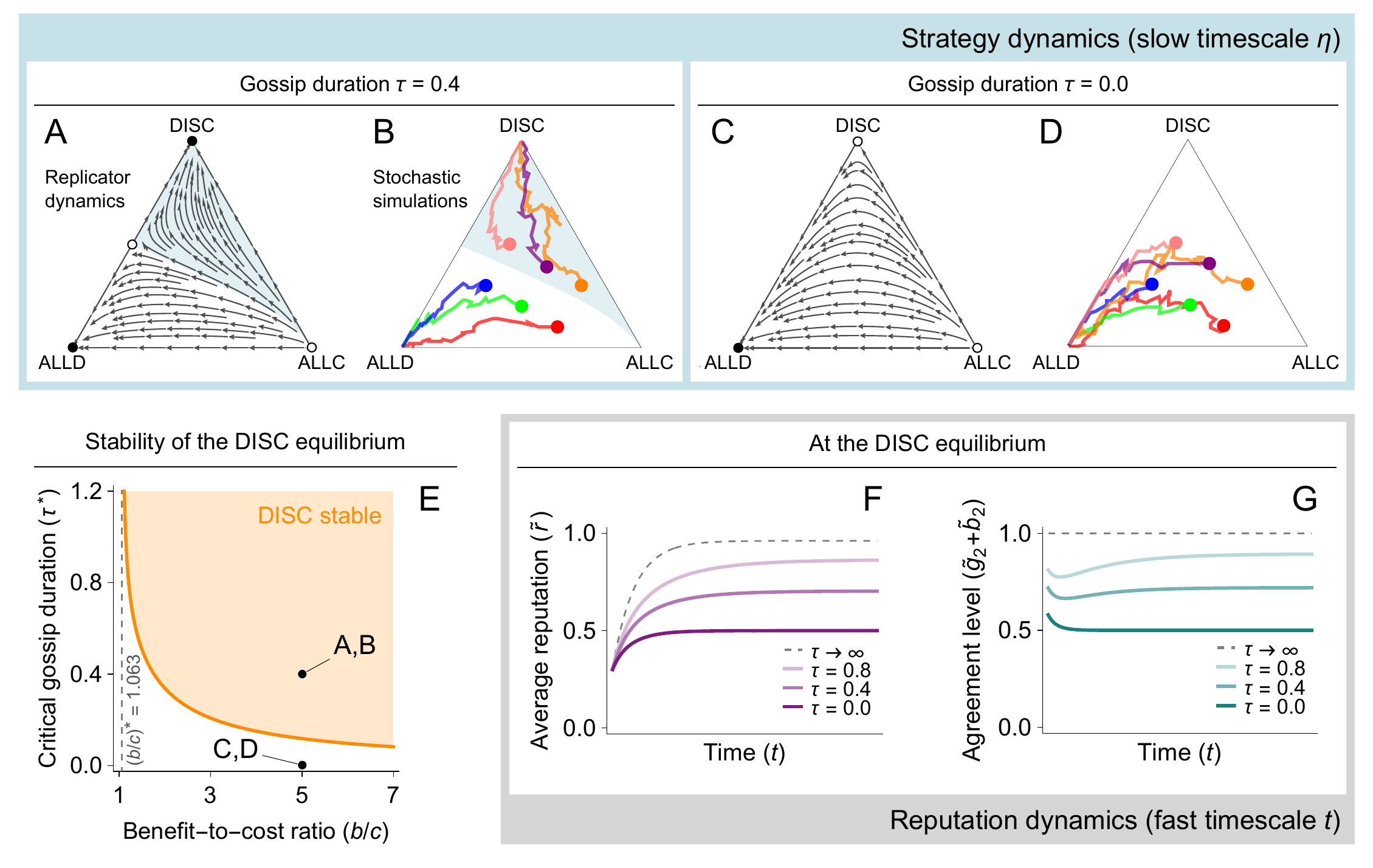}
    \caption{
    \textbf{Sufficiently long gossip stabilizes cooperation.} 
    \textbf{A--D}: Dynamics of competition among strategies $\allc$, $\alld$, and $\disc$ under the Stern Judging norm, with a fixed benefit-to-cost ratio ($b/c=5$). 
    \textbf{A, C}: Gradients of selection (arrows) in the replicator dynamics (\cref{eq:replicator}).
    There is a basin of attraction towards the $\disc$ vertex (shaded region) when $\sgen=0.4>\sgen^*$ (A) but not when $\sgen=0.0<\sgen^*$ (C).
    \textbf{B, D}: Trajectories of stochastic simulations in a finite population ($N\!=\!100$), with colors indicating different initial conditions (see \nameref{sec:methods}).
    The long-term behavior of the stochastic simulations is consistent with the analytical predictions of the replicator-dynamic ODEs (B vs A; D vs C): When $\sgen=0.4>\sgen^*$ (B), trajectories starting from initial conditions above the separatrix tend to converge to the $\disc$ vertex (shaded region denotes the basin of attraction in A). When $\sgen=0.0<\sgen^*$ (D), all six trajectories converge to the $\alld$ vertex.
    \textbf{E}: The discriminator-only equilibrium ($f_\alld=1$) is locally stable only if the scaled gossip duration $\sgen$ exceeds a critical value $\sgen^*$ (solid orange line defined by condition ii) and $b/c$ exceeds a critical value $(b/c)^*$ (dashed gray line defined by condition i). The orange region indicates parameter values where both these conditions are satisfied. The critical gossip duration $\sgen^*$ decreases with the benefit-to-cost ratio, $b/c$.
    \textbf{F, G}: The average reputation and agreement level as a function of time $t$ during the process of reputation dynamics by independent observations and gossip (\cref{eq:repODE}). These quantities are evaluated at the $\disc$ vertex ($f_\disc=1$) with a fixed benefit-to-cost ratio ($b/c=5$). Colors correspond to different values for the duration of gossip periods, $\sgen$. The darkest color in each panel corresponds to $\sgen=0.0$ (no gossip), which is equivalent to private reputations. The dashed lines correspond to $\sgen\rightarrow\infty$ (infinitely long gossip), which is equivalent to public reputations.
    Other parameters: $\assess=\exec=0.02$.
    Analogous results for the Simple Standing and Shunning norms are shown in \cref{fig:stabilityALL} (see also \nameref{sec:methods}).
    }
    \label{fig:stabilitySJ}
\end{figure}

When reputations are assessed privately without gossip, competition among cooperators ($\allc$), defectors ($\alld$), and discriminators ($\disc$) will always lead to a population of pure defectors under the Stern Judging norm. That is, the only stable strategic equilibrium is $f_\alld=1$, regardless of the benefits and costs of cooperation ($b$ and $c$) or the rates of erroneous action or assessment ($\exec$ and $\assess$) \cite{okada_solution_2018}. Gossip can qualitatively change this outcome. For example, when gossip occurs for duration $\sgen=0.4$ (\cref{fig:stabilitySJ}A), both $\alld$ and $\disc$ are stable strategic equilibria; indeed, there is a large basin of initial conditions that lead the population to the $\disc$ equilibrium, which supports high levels of cooperation. This basin disappears in the absence of gossip ($\sgen=0.0$; \cref{fig:stabilitySJ}C). Stochastic simulations in finite populations show agreement with these analytical predictions derived from the infinite-population replicator-dynamic ODEs (\cref{eq:replicator}; \cref{fig:stabilitySJ}B, D). Thus, at least under Stern Judging, gossip can sometimes stabilize cooperation. 

The key question remains: how much gossip is required to sustain cooperation? Is an arbitrarily small but positive amount of gossip sufficient? To answer these questions, we derive an analytical condition for the stability of the discriminator equilibrium. First, to compute the equilibrium reputations, we substitute the agreement and disagreement terms from the peer-to-peer gossip process (\cref{eq:g2b2d2-wf-largeN}) into the expressions in \cref{eq:rep_ps}, and we set the right-hand side of the fast-time reputation ODEs (\cref{eq:repODE}) to zero. These equilibrium reputations in turn determine the payoffs to strategic types (\cref{eq:payoffs}), which we substitute into the replicator-dynamic ODEs (\cref{eq:replicator}). Linear stability analysis (\nameref{sec:methods-linear} in \nameref{sec:methods}; see also \SI) shows that the discriminator equilibrium ($f_\disc=1$) is locally stable under the Stern Judging norm if and only if the following conditions are both satisfied:
\begin{align}
    & \textrm{(i)  }\quad \dfrac{b}{c} > \left(\dfrac{b}{c}\right)^* 
    = \dfrac{1}{\left(1-2 \assess\right) (1-\exec)}
    \label{eq:cond1SJ}
    \qquad\textrm{and}\qquad
    \\
    & \textrm{(ii)  }\quad \sgen
    > \sgen^*
    = \log\left[  \left(2 -  \dfrac{\left(\frac{b}{c}\right)}{\left(\frac{b}{c}\right)-\frac{1}{2 (1-\assess)}}\right)
    \left(\dfrac{\left(\frac{b}{c}\right)}{\left(\frac{b}{c}\right)-
    \left(\frac{b}{c}\right)^{^*}
    }\right)\right] \;.
    \label{eq:cond2SJ}
\end{align}
The first condition above is identical to the minimum benefit-to-cost ratio $(b/c)^*$ required to stabilize the discriminator equilibrium under fully public information (dashed line in \cref{fig:stabilitySJ}E), which is already known in the literature \cite{kessinger_evolution_2023}. The second condition gives, in addition, the critical gossip duration $\sgen^*$ required to stabilize cooperation. Note that $\sgen^*$ is a decreasing function of the benefit-to-cost ratio $b/c$ ($\partial \sgen^* / \partial (b/c) < 0$)---which means that less gossip is required to stabilize cooperation when the benefits of mutual cooperation are greater (\cref{fig:stabilitySJ}E). The duration of gossip required $\sgen^*$ approaches infinity as $b/c\rightarrow(b/c)^*$, meaning that no amount of gossip can outperform fully public information (at least when there is no bias in gossip transmission, an assumption we will later relax). Conversely, $\sgen^*$ approaches zero as $b/c \rightarrow \infty$, which means that a positive amount of gossip is always required to stabilize cooperation, except in the limit of an infinite benefit-to-cost ratio.

Gossip stabilizes cooperation because it increases agreement about reputations---even in the presence of errors---and consequently improves how discriminators view each other on average. To demonstrate this, we plot the average reputation (\cref{fig:stabilitySJ}F) and average agreement level (\cref{fig:stabilitySJ}G) in the population at the discriminator-only equilibrium ($f_{\disc}=1$) as a function of time $t$ for different durations of gossip $\sgen$. In the absence of gossip ($\sgen=0$), both quantities are $1/2$, in agreement with results under fully private information \cite{radzvilavicius_evolution_2019}. As the gossip duration $\sgen$ increases, both agreement and average reputation increase. In the limit of infinitely long gossip ($\sgen\rightarrow\infty$), we achieve the same average reputation and agreement level as under fully public information (dashed lines in \cref{fig:stabilitySJ}F and G). In this sense, our model of gossip spans the spectrum from fully private to fully public information about social reputations.

Conditions (i) and (ii) also reveal how errors modulate the effects of gossip. Since errors in either reputation assessment or strategy execution increase the possibility of misassigned reputations and therefore disagreement, we might expect that gossip would need to proceed for longer to counteract their destabilizing effects. Indeed, we can prove that $\sgen^*$ is monotonically increasing with the error rates: $\partial \sgen^*/\partial \assess>0$ and $\partial \sgen^*/\partial \exec>0$ (see \SI\ Section~\ref{sec:si-errors} and \cref{fig:errorsSJ}).

\subsection{Noisy gossip is less beneficial for cooperation} \label{sec:result3}

We have assumed that reputation information is transmitted faithfully during peer-to-peer gossip. However, in reality, gossip transmission is a noisy and possibly even biased process, just like in the game of telephone: an individual might hear from a source that a focal individual is good, but that individual might convey the opposite information to the next individual in line, either accidentally (e.g., misunderstanding) or intentionally (e.g., preferential treatment or malice).

To account for noise in transmission, we introduce the possibility of ``mutation" in the Wright-Fisher process describing peer-to-peer gossip over subsequent rounds (or ``generations''). Suppose that, in round $\gen$, there are $\ell$ individuals who believe a given focal individual $i$ is good and $N-\ell$ who believe individual $i$ is bad. We now assume that  an individual who consults a peer who believes $i$ is good will, with probability $\GtoB$, adopt the opposite opinion (``mutate'') in round $\gen+1$. Likewise, an individual who consults a peer who believes $i$ is bad will, with probability $\BtoG$, adopt the opposite opinion. In the absence of mutation ($\GtoB=\BtoG=0$), we recover the model of noiseless gossip.

We let $R_{i,\gen}\in\{0,1/N,\dots,\left(N-1\right)/N,1\}$ be a random variable that tracks the frequency of individuals who view individual $i$ as good in round $\gen$. Assuming that (1) $N$ is large, (2) $\GtoB$ and $\BtoG$ are small, and (3) a fraction $r_{i,0}$ view $i$ as good at the start of gossip ($\gen=0$), we can approximate the mean and variance after $\gen$ generations of peer-to-peer gossip (equivalent to duration $\sgen\triangleq\gen/N$, as before) about focal individual $i$ as 
\begin{equation}\small
    \label{eq:gossip}
    \begin{split}
        \E\left[R_{i,\sgen} \big| R_{i,0}=r_{i,0} \right] 
        & = \left(r_{i,0}-\dfrac{\BTOG}{\GTOB+\BTOG}\right) e^{-\left(\GTOB+\BTOG\right)\sgen} + \dfrac{\BTOG}{\GTOB+\BTOG} 
        \;,
        \\
        \Var \left(R_{i,\sgen} \big| R_{i,0}=r_{i,0} \right)
        & = \dfrac{\BTOG}{\GTOB+\BTOG} \left(1-\dfrac{\BTOG}{\GTOB+\BTOG}\right) 
        \cdot
        \dfrac{1}{1 + 2\left(\GTOB+\BTOG\right) } \left(1 - e^{-\left(2\left(\GTOB+\BTOG\right)+1\right)\sgen}\right)
        \\ & \quad + \left(1-\dfrac{2\BTOG}{\GTOB+\BTOG}\right) \left(r_{i,0}-\dfrac{\BTOG}{\GTOB+\BTOG}\right) \dfrac{1}{ 1 + \left(\GTOB+\BTOG\right)}
        \cdot 
        e^{-\left(\GTOB+\BTOG\right)\sgen} \left(1 - e^{-\left(\left(\GTOB+\BTOG\right)+1\right)\sgen}\right)
        \\ & \quad - \left(r_{i,0}-\dfrac{\BTOG}{\GTOB+\BTOG}\right)^2 e^{-2\left(\GTOB+\BTOG\right)\sgen} \left( 1 - e^{-\sgen} \right)
        \;,
    \end{split}
\end{equation}
where $\GTOB=N\GtoB$ and $\BTOG=N\BtoG$ are scaled mutation rates \cite[\SI;][]{tataru_inference_2015,tataru_statistical_2017}. 

As in the case of noiseless gossip described earlier, we assume that the gossip occurs independently for each focal individual and that the fraction of the population who view a focal individual $i$ of type $\strat$ as good at the start of a gossip period is equal the fraction of the population who view type $\strat$ as good in the context of reputation ODEs in \cref{eq:repODE} (i.e., if individual $i$ is of type $\strat$, then $r_{i,0}=r_\strat$). Agreement and disagreement terms after a period of gossip of duration $\sgen$ can then be computed as
\begin{equation}
    \label{eq:g2b2d2-wf-bias}
    \begin{split} 
        \gtwo
        & = \sum_\strat f_\strat \cdot \E\left[R_{i,\sgen}^2 \big| R_{i,0}=r_\strat \right] 
        \;,
        \\
        \btwo 
        & = \sum_\strat f_\strat \cdot \E\left[(1-R_{i,\sgen})^2 \big| R_{i,0}=r_\strat \right]
        \;,
        \\
        \dtwo 
        & = \sum_\strat f_\strat \cdot \E\left[R_{i,\sgen}(1-R_{i,\sgen}) \big| R_{i,0}=r_\strat \right]
        \;,
    \end{split}
\end{equation}
where $\E\big[R_{i,\sgen}^2 | R_{i,0}=r_\strat \big]$, $\E\big[(1-R_{i,\sgen})^2 | R_{i,0}=r_\strat \big]$, and $\E\big[R_{i,\sgen}(1-R_{i,\sgen}) | R_{i,0}=r_\strat \big]$ can be expressed in terms of the mean and variance of $R_{i,\gen}$ (\cref{eq:gossip}; \SI).

Importantly, in the case of noisy transmission, gossip affects not only the variance but also the mean proportion of the population who view a focal individual as good (\cref{eq:gossip}). To account for this, we must replace the expressions for $p_\strat$ (\cref{eq:rep_ps}), the probability that an individual of strategic type $\strat$ earns a good reputation, with the following:
\begin{equation}
    \begin{split}
        p_{\allc}(t) & = \rtil(t) \pgc + \left(1-\rtil(t)\right) \pbc \;, \\
        p_{\alld}(t) & = \rtil(t) \pgd + \left(1-\rtil(t)\right) \pbd \;, \\
        p_{\disc}(t) & = \gtwo(t) \pgc + \dtwo(t) \left(\pbc + \pgd\right) + \btwo(t) \pbd \;,
    \end{split}
\end{equation}
where $\rtil$ is the average reputation in the population after gossip of duration $\sgen$:
\begin{equation}
    \rtil = \sum_\strat f_\strat \cdot \E\left[R_{i,\sgen} | R_{i,0}=r_\strat \right] 
    = \left(r -\frac{\BTOG}{\GTOB+\BTOG}\right) e^{-\left(\GTOB+\BTOG\right)\sgen} + \frac{\BTOG}{\GTOB+\BTOG} \;.
\end{equation}
We recover the case of noiseless gossip (\cref{eq:rep_ps}) by letting $\GTOB=\BTOG=0$ and setting $0/0:=1$ \cite{tataru_statistical_2017}; in particular, in the absence of noise, gossip does not affect the average reputation in the population ($\rtil=r$).

\begin{SCfigure}[1.6][h!]
    \includegraphics[width=0.44\linewidth]{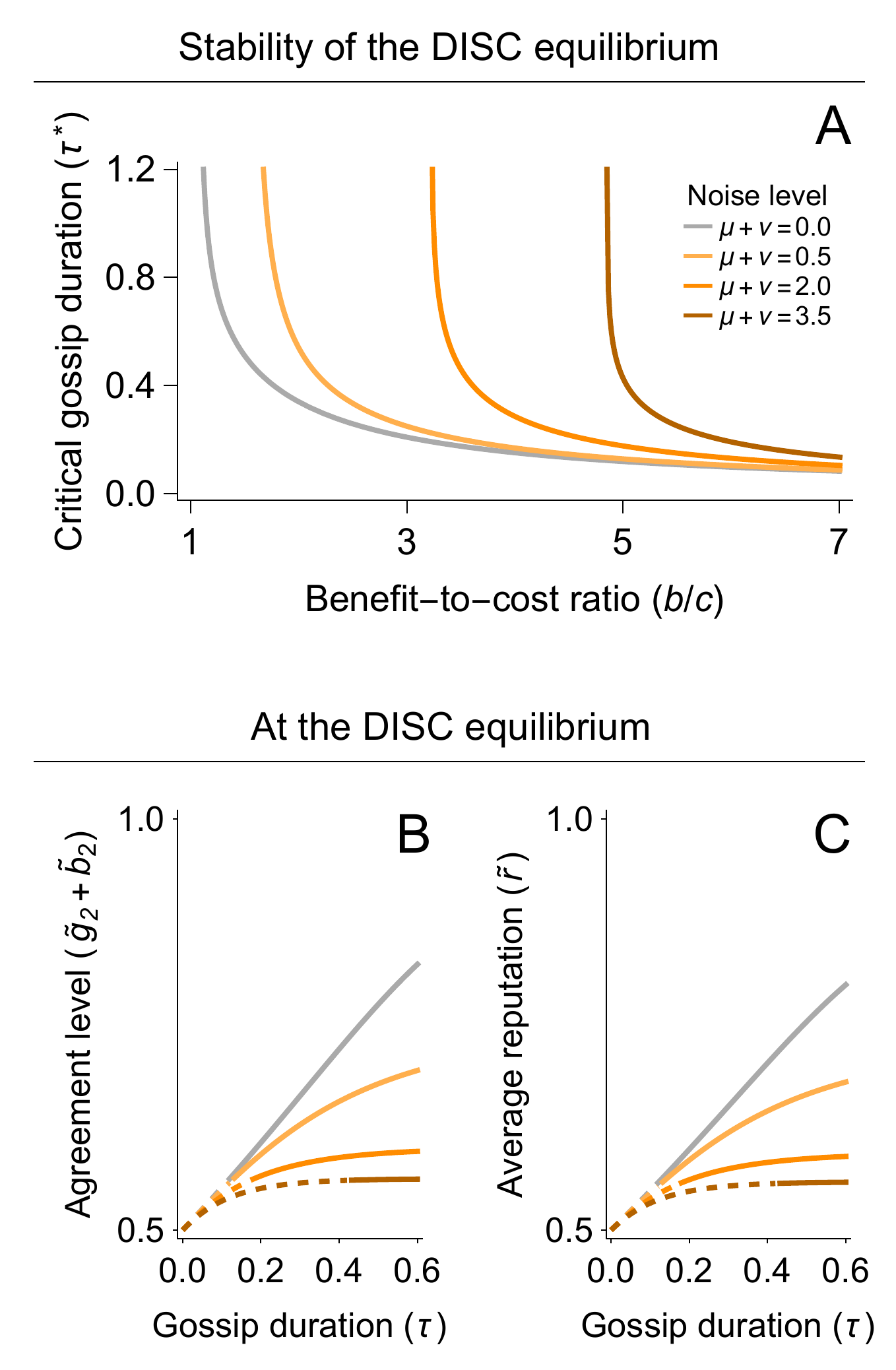}
    \caption{
    \textbf{Noise in gossip transmission tends to destabilize cooperation.}
    \textbf{A}: The critical gossip duration $\sgen^*$ required to stabilize cooperation as a function of the benefit-to-cost ratio $b/c$, under the Stern Judging norm. 
    Colors denote results for different amounts of unbiased noise in gossip ($\GTOB+\BTOG$). The gray line indicates the critical gossip duration for noiseless transmission ($\GTOB\!=\!\BTOG\!=\!0$; \cref{fig:stabilitySJ}E).
    \textbf{B, C}: The equilibrium average reputation and agreement level at the $\disc$ vertex as a function of scaled gossip duration $\sgen$, evaluated with a fixed benefit-to-cost ratio ($b/c=5$).  
    Colors are as indicated in A. Solid (dashed) segments denote parameters for which the $\disc$-only equilibrium is locally stable (unstable).
    Other parameters: $\assess=\exec=0.02$.
    Analogous results for the Simple Standing and Shunning norms are shown in \cref{fig:noiseALL}.
    }
    \label{fig:noiseSJ}
\end{SCfigure}

Noise in gossip transmission makes it more difficult to stabilize cooperation (\cref{fig:noiseSJ}). Under the Stern Judging norm and for a given benefit-to-cost ratio $b/c$, the duration of gossip $\sgen^*$ required to stabilize cooperation increases with the amount of noise ($\GTOB+\BTOG$), even when there is no bias in transmission ($\GTOB=\BTOG$; \cref{fig:noiseSJ}A). In other words, as gossip becomes more prone to noise in transmission, the population must engage in gossip for longer in order to stabilize the all-$\disc$ equilibrium. This is because transmission noise, much like errors in assessment or execution, decreases the level of agreement in the population (\cref{fig:noiseSJ}C) and, consequently, decreases the average reputation of discriminators (\cref{fig:noiseSJ}B). Importantly, noisy gossip hinders cooperation even in the limit that otherwise corresponds to public information: the higher the level of noise, the higher the minimum benefit-to-cost ratio $b/c$ required to sustain cooperation in the limit of infinitely long gossip (i.e., the vertical asymptotes in \cref{fig:noiseSJ}A).

\subsection{Biased gossip can facilitate cooperation}

In real-world scenarios, gossip is not only noisy, but it may also be biased: Someone who directly judges a focal individual as good or learns this through gossip may nonetheless report the individual as bad in a subsequent round of peer-to-peer gossip. Or, conversely, gossip may be biased towards reporting bad individuals as good. Biases may arise either by mistake (such as a cognitive bias towards a pessimistic or optimistic view of people's reputations) or by design (such as malice or forgiveness). In either case, we wish to understand how biased gossip affects reputations in a population and, in turn, modifies the stability of cooperation.

To study the effects of bias, we fix the total magnitude of noise in transmission ($\GTOB+\BTOG$), and we compute the critical gossip duration $\sgen^*$ required to stabilize cooperation (at $f_\disc=1$) as a function of the \textit{gossip bias} $\beta  \triangleq 2\big(\frac{\BTOG}{\GTOB+\BTOG}-\frac{1}{2}\big) \in[-1,1]$. Here $\beta=-1$ indicates maximally negative bias (i.e., any noise in gossip transmits a positive reputation as a negative reputation), $\beta=+1$ indicates maximally positive bias, and $\beta=0$ indicates no bias.

\begin{figure}[h!]
    \centering
    \includegraphics[width=0.91\linewidth]{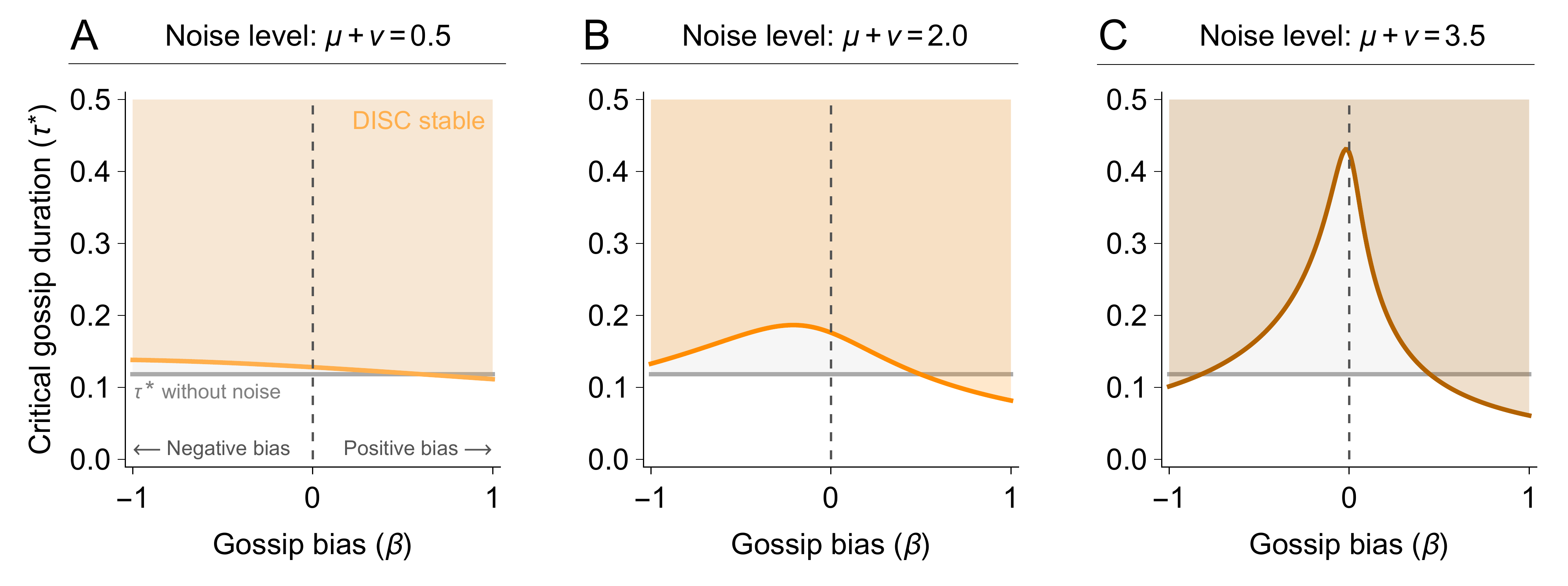}
    \caption{
    \textbf{Biased gossip can facilitate or impede cooperation.}
    Critical gossip duration $\sgen^*$ required for stable cooperation (solid orange line) as a function of the strength of gossip bias $\beta$, under the Stern Judging norm. 
    The shades of orange denote three different amounts of noise, as in \cref{fig:noiseSJ}: $\GTOB+\BTOG=0.5$ (A), $\GTOB+\BTOG=2.0$ (B), and $\GTOB+\BTOG=3.5$ (C).
    Orange regions indicate parameter regimes where the $\disc$-only equilibrium is locally stable. Solid gray lines (identical across panels) indicate the baseline critical gossip duration $\sgen^*$ in the absence of transmission noise ($\GTOB=\BTOG=0$; see \cref{fig:stabilitySJ}E).
    Other parameters: $b/c=5$, $\assess=\exec=0.02$.
    Analogous results for the Simple Standing and Shunning norms are shown in \cref{fig:biasALL}.
    }
    \label{fig:biasSJ}
\end{figure}

Under the Stern Judging norm, biased gossip has asymmetric and sometimes even non-monotonic effects on the duration of gossip required for cooperation (\cref{fig:biasSJ}). When the overall amount of transmission noise is small ($\GTOB+\BTOG=0.5$, \cref{fig:biasSJ}A), the critical gossip duration $\sgen^*$ decreases monotonically with bias: the more positive the bias, the less gossip is required to stabilize the cooperation, as individuals converge towards more positive views of each other and are more likely to cooperation. But the effect of bias becomes non-monotonic when gossip is more noisy ($\GTOB+\BTOG=2.0$, \cref{fig:biasSJ}B or $\GTOB+\BTOG=3.5$, \cref{fig:biasSJ}C). Positive bias ($\beta>0$) is still increasingly beneficial for cooperation with increasing magnitude; however, a small amount of negative bias is detrimental to cooperation, whereas a large negative bias is actually beneficial for cooperation. The basic intuition is that large amounts of noise cause disagreement and tend to destabilize cooperation for a given duration of gossip, but this effect can be counterbalanced by either positive bias or strong negative bias.

We can understand these patterns in terms of the effects of bias on agreement and disagreement levels (\cref{fig:agreementSJ}). When noise is rare ($\GTOB+\BTOG=0.5$), the quantity $g_2$ increases monotonically with the transmission bias $\beta$ whereas the quantities $b_2$ and $d_2$ decrease monotonically (compare \cref{fig:agreementSJ}A--C with \cref{fig:biasSJ}A). But when gossip is more noisy ($\GTOB+\BTOG=2.0, 3.5$), all three quantities $g_2$, $b_2$, and $d_2$ are non-monotonic in $\beta$ (\cref{fig:agreementSJ}D--F and \cref{fig:agreementSJ}G--I), producing a non-monotonic effect of bias on the stability of cooperation (\cref{fig:biasSJ}B--C).

Whereas unbiased noise tends to destabilize cooperation (\cref{fig:noiseSJ}), biased noisy transmission can expand the region of stable cooperation, compared even to the case of no noise. In other words, the critical gossip duration $\sgen^*$ required for stable cooperation may be shorter for noisy transmission with a strong bias than for noiseless gossip (\cref{fig:schematic}; solid gray lines in \cref{fig:biasSJ}B,C). This is because when individuals overwhelmingly tend to transmit either positive (or even negative) gossip, the population will come to an agreement more quickly than if positive and negative noise are equally likely. 

We have also analyzed the impact of biased gossip under the Simple Standing and Shunning norms (\cref{fig:biasALL}). Unlike the case of Stern Judging, for Simple Standing (\cref{fig:biasALL}A--C) and Shunning (\cref{fig:biasALL}D--F), the duration of gossip $\sgen^*_\alld$ required to stabilize $\disc$ against $\alld$ decreases monotonically with gossip bias $\beta$. This monotonicity reflects the fact that, unlike Stern Judging, Simple Standing and Shunning do not distinguish between justified and unjustified behavior toward individuals with bad reputations. Thus, for these two norms, increasing the magnitude of negative bias increases the frequency of bad reputations in the population, and cooperation becomes more difficult to sustain (i.e., $\sgen^*_\alld$ increases).

\section{Discussion}

We have developed a mechanistic model of gossip about social reputations and studied the effects of gossip on cooperative behaviors. In our analysis, individuals privately assess each other's reputations, and then they modify their views either by consulting a designated source of gossip or, equivalently under a transformation of parameters, by exchanging views with randomly selected peers. Iterative rounds of private observation and gossip eventually produce equilibrium reputations that determine the payoffs achieved by three different behavioral strategies. Individuals can then imitate each other's strategies on a slower timescale, which may lead to long-term stable cooperation. This integrated model of gossip and indirect reciprocity spans continuously between the classical cases of fully private and fully public information. This approach allows us to analyze how the quantity and quality of gossip transmission affect long-term behavior and collective welfare.

We have shown that sufficiently long periods of gossip can stabilize cooperation. Gossip increases agreement about reputations, even in the presence of erroneous actions and assessments. The increased agreement, in turn, reduces the likelihood that a cooperative action is judged as unjustified behavior, and it improves the reputations in the population as a whole. In other words, even when reputations are assessed privately without any top-down public institution to enforce agreement, a bottom-up process based on peer-to-peer gossip can build consensus in the population, and if gossip periods are sufficiently long, stable cooperation can be restored. In this sense, our model offers a mechanistic justification for the common assertion that gossip can facilitate cooperation by indirect reciprocity \cite{nowak_evolution_2005,ohtsuki_indirect_2009,santos_social_2018,okada_tolerant_2017,okada_solution_2018,radzvilavicius_evolution_2019,radzvilavicius_adherence_2021,harrison_strength_2011,kessinger_evolution_2023,morsky_indirect_2023}. Whereas prior work has explored gossip-based cooperation using agent-based simulations \cite{ohtsuki_indirect_2009,righi_gossip_2022}, our mean-field analysis provides an analytical expression for the minimum amount of gossip required to sustain cooperation---which allows us to understand how errors in action and assessment, as well as cooperative benefits and costs, govern the amount of gossip necessary for cooperation.

A key insight from our analysis is that peer-to-peer gossip stimulates consensus about reputations only if it occurs in a finite population. Indeed, if the population size were infinite, then no (finite) amount of gossip could ever change the level of agreement. We can understand this insight mathematically in \cref{eq:g2b2d2-wf}, where, regardless of the gossip duration $\gen<\infty$, the levels of agreement before and after gossip are identical in the infinite-population limit ($\lim_{N\rightarrow\infty} \gtwo = g_2$). Because of this, our model accounts for a finite population when describing the dynamics of gossip, whereas it describes the dynamics of strategic changes (on a slower timescale) in an infinite population. This mixture of finite- and infinite-population treatments mirrors models of indirect reciprocity that track reputations using finite `image matrices' while tracking strategy frequencies using replicator dynamic ODEs \cite{uchida_effect_2010,uchida_effect_2013}. We have confirmed that the predictions of our analytical treatment based on this approach are consistent with the behavior of the corresponding discrete stochastic simulations in finite populations (\cref{fig:stabilitySJ}).

Our account of how gossip facilitates cooperation by fostering consensus about reputations complements a larger literature on how gossip facilitates partner choice. The theory of reputation-based partner choice posits that when individuals must compete for interaction partners, players are motivated to cooperate more because those with good reputations tend to attract more cooperative partners in the future \cite{wu_when_2015,feinberg_gossip_2014,traag_indirect_2011,traag_dynamical_2013}. This theory hinges on the ability of individuals to alter their social ties, whereas our analysis shows that gossip can promote cooperative behavior even when the social environment is fixed and everyone must interact with everyone else.

Our results raise several unresolved questions about the role of population structure in gossip-based cooperation. Our single-source gossip model, as the name suggests, assumes that only one individual initiates gossip. In reality, however, multiple sources within a population may spread different and potentially conflicting information about others. Future research could explore how the number of gossip sources and the algorithm by which receivers integrate received information impacts cooperation. In addition, our peer-to-peer model assumes that players can exchange gossip with anyone (i.e., a complete gossip network), and as a result the population approaches full consensus as gossip duration increases. But under what conditions (e.g., modular or dynamic gossip networks) would the population split into different camps that view a focal individual differently, and how would polarization in reputations impact cooperation? How does the number of sub-groups in a population---as well as their relative sizes and internal structures---affect the rate of convergence to consensus and stability of cooperation, and how do differential rates of gossip within and between sub-groups modulate these effects? These questions remain open for future research.

One limitation of gossip is that reputations are not always transmitted faithfully \cite{seki_model_2016,wu_honesty_2021,dorescruz_gossip_2021}. Noise during transmission can either be unbiased (e.g., accidental errors) or biased (e.g., intentional misrepresentation). We have shown that unbiased noise tends to undercut the benefits of gossip. This result is perhaps unsurprising because, much like errors in assessment or execution \cite{hilbe_indirect_2018}, transmission noise impedes agreement in the population. It is notable, however, that biased gossip can sometimes stabilize cooperation relative to unbiased gossip or relative even to noiseless gossip. This is true for all norms studied when gossip is biased toward positive reports. In addition, under the Stern Judging norm, a strong negative bias can also stabilize cooperation, compared to noiseless gossip. While this is potentially good news---cooperation does not necessarily unravel even when gossip is biased---our analysis has been limited to populations with uniform tendencies to transmit false or manipulated information. An important extension for future research is to study heterogeneity in how bias is applied. For example, in a population with group structure, individuals may have different levels of bias when gossiping about in-group versus out-group members (e.g., $\beta_{\text{in-group}}>0$ and $\beta_{\text{out-group}}<0$).

We have assumed that the propensity to gossip is both uniform across the population and exogenously fixed. But competition between gossip strategies may complicate the picture: for example, previous work has found that dishonest gossip strategies---where gossipers deterministically transmit false information (i.e., pure bias, akin to $\GtoB=1$ or $\BtoG=1$)---can outperform honest gossip strategies under certain conditions, although dishonest gossip tends to undermine cooperation \cite{wu_gossip_2016,nakamaru_evolution_2004,seki_model_2016}. A natural question, then, is whether the amount and quality of gossip that stabilizes cooperation would naturally evolve if individuals were allowed to adjust the frequency and fidelity with which they transmit information. A recent study on the role of empathy in indirect reciprocity has found that populations can evolve empathetic evaluation under the right conditions \cite{radzvilavicius_evolution_2019}---which implies that single-source gossip ($\queen>0$) can also evolve under the same conditions. However, several questions about the evolution of peer-to-peer gossip remain unresolved. If the degrees of noise and bias are subject to selection, will individuals evolve to share information truthfully and accurately? And how will the long-term dynamics differ when gossip transmission is inherently costly? Future research on the evolution of gossip strategies may guide the design of incentives for individuals to adopt gossip behaviors that ultimately promote collective welfare. 

Communication has long been recognized as a key factor in human cooperation \cite{ostrom_communication_1991,dunbar_gossip_2004}. And there is already an extensive literature on how opinions spread in a population through peer-to-peer communication, including the complications of explicit population structure and complex modes of contagion \cite{baronchelli2018emergence,centola2018experimental,centola2018behavior,vasconcelos2019consensus}. Despite this, only recently have researchers begun to develop mathematically tractable frameworks to study strategic evolution coincident with opinion spread \cite[e.g.,][]{seki_model_2016,zhang_conformity_2023}. Our study provides a minimal, mechanistic description of how gossip facilitates consensus about reputations---a critical component of cooperation by indirect reciprocity. There remains a large, uncharted realm of research that combines the complex dynamics of belief contagion with the dynamics of social behaviors conditioned on individuals' beliefs.

\vspace{5ex}

\newpage
\section{Materials and Methods}
\label{sec:methods}

\phantomsection

Here we provide additional details of our mathematical model (\nameref{sec:model}). We refer the reader to \SI\ for detailed derivations.

\subsection{Reputation dynamics} \label{sec:methods-repODE}

Let $\delta$ be the observation rate; $\delta\Delta t$ is the probability that an observer $j$ observes a given focal individual $i$ in an interval $\Delta t$. 
If $i$'s reputation in the eyes of $j$ at time $t$ is $r_{ij}(t)$, then the expected reputation of $i$ in the eyes of $j$ after $\Delta t$ is
\begin{equation}
    r_{ij}(t+\Delta t) 
    = \underbrace{\left(1-\delta \Delta t\right)}_{\mathbb{P}(j \textrm{ does not observe } i)} \cdot\ r_{ij}(t) 
    + \underbrace{\left(\delta \Delta t\right)}_{\mathbb{P}(j \textrm{ observes } i)} \cdot \underbrace{p_{ij} (t)}_{\mathbb{P}(j \textrm{ views } i \textrm{ as good})} \;.
\end{equation}
Assuming that a single round of observations takes place for every round of interactions ($\delta=1$), we have
\begin{equation}
    \dfrac{dr_{ij}}{dt} = \lim_{\Delta t\rightarrow 0} \dfrac{r_{ij}(t+\Delta t) - r_{ij}(t)}{\Delta t} = p_{ij}(t) - r_{ij}(t)\;.
\end{equation}
We also assume that, in a given observation round, observers observe an independently and randomly selected interaction of each donor. Under this assumption, the dynamics of the average reputation of each strategic type $\strat$ follow to the ODEs \cite[][]{perret_evolution_2021},
\begin{equation}
    \dfrac{dr_\strat}{dt} = p_\strat(t) - r_\strat(t) \;,
\end{equation}
as reported in \cref{eq:repODE}.

\subsection{Social norms} \label{sec:methods-norms}

A social norm is a set of assessment rules used to assign reputations. A norm is considered `first-order' if it updates a donor's reputation based solely on the action of the donor, and `second-order' if it uses both the donor's action and the recipient's reputation to assess the donor. We focus on second-order norms because more complex norms, while possible, while more complex norms are possible, they typically produce less cooperation than these simple norms \cite{santos_social_2018}.

We consider three second-order social norms that are most common in studies of indirect reciprocity \cite{sasaki_evolution_2017,radzvilavicius_evolution_2019,radzvilavicius_adherence_2021,kessinger_evolution_2023}: Stern Judging {\tiny$\setlength\arraycolsep{1pt}\begin{pmatrix} G & B \\ B & G \end{pmatrix}$}, Simple Standing {\tiny$\setlength\arraycolsep{1pt}\begin{pmatrix} G & G \\ B & G \end{pmatrix}$}, and Shunning {\tiny$\setlength\arraycolsep{1pt}\begin{pmatrix} G & B \\ B & B \end{pmatrix}$}. In each binary matrix, the rows indicate the donor's action (row one for cooperation, two for defection), the columns indicate the recipient's reputation (column one for good, two for bad), and the entries indicate how the donor is assessed ($G$ for good, $B$ for bad) under the corresponding norm \cite{radzvilavicius_evolution_2019}.

Each of the three norms can be parametrized as $(\p,\q)$, where the parameter $\p$ ($\q$) denotes the probability that cooperating with (defecting against) a bad recipient yields a good standing. We have $(\p,\q)=(0,1)$ for Stern Judging, $(1,1)$ for Simple Standing, and $(0,0)$ for Shunning.

\subsection{Reputation assessments} \label{sec:methods-ps}

Next, we derive the probability $p_\strat$ that an individual of strategic type $\strat$ earns a good reputation after gossip (\cref{eq:rep_ps}), following the approach described in \citet{radzvilavicius_evolution_2019,radzvilavicius_adherence_2021} and \citet{kessinger_evolution_2023}. 

Recall that $\rtil$ is the post-gossip average reputation in the population; $\gtwo$ ($\btwo$) is the post-gossip probability that two randomly selected individuals agree that a third individual is good (bad); and $\dtwo$ is the post-gossip probability that the first thinks the third is good but the second thinks the third is bad.
For convenience, we also define the following quantities \cite{kessinger_evolution_2023}:
\begin{equation}
    \label{eq:pgcetc}
    \begin{split}
        \pgc & = \left(1-\exec\right)\left(1-\assess\right) + \exec \assess \triangleq \varepsilon \;, \\
        \pgd & = \assess\;, \\
        \pbc & = \p \left(\varepsilon - \assess\right) + \q \left(1 - \varepsilon - \assess\right) + \assess\;,\\
        \pbd & = \q \left(1 - 2 \assess\right) + \assess\;,
    \end{split}
\end{equation}
where $P_{XY}$ is the probability that a donor who intends to $Y\in$\{cooperate ($C$), defect ($D$)\} with a recipient viewed as $X\in$\{good ($G$), bad ($B$)\} by the observer is assigned a good reputation.

\textbf{Cooperators ($\allc$).} 
A cooperator ($\allc$) gains a good reputation by either 
(1) interacting with someone with a good reputation (with probability $\rtil$), intending to cooperate, and successfully being assigned a good reputation (with probability $\pgc$); or 
(2) interacting with someone with a bad reputation (with probability $1-\rtil$), intending to cooperate, and erroneously being assigned a good reputation (with probability $\pbc$).

Thus, the probability that a cooperator earns a good reputation is given by
\begin{equation}
    p_\allc = \rtil \pgc + (1 - \rtil) \pbc \;.
\end{equation}

\textbf{Defectors ($\alld$).} 
Similarly, a defector ($\alld$) gains a good reputation by either
(1) interacting with someone with a good reputation (with probability $\rtil$), intending to defect, and erroneously being assigned a good reputation (with probability $\pgd$); or
(2) interacting with someone with a bad reputation (with probability $1-\rtil$), intending to defect, and successfully being assigned a good reputation (with probability $\pbd$).

Thus, the probability that a defect earns a good reputation is given by
\begin{equation}
    p_\alld = \rtil \pgd + (1 - \rtil) \pbd \;.
\end{equation}

\textbf{Discriminators ($\disc$).}
Finally, a discriminator ($\disc$) gains a good reputation by
\begin{itemize}
    \item[(1)] interacting with someone who has a good reputation in the eyes of both the donor and the observer (with probability $\gtwo$), intending to cooperate, and being assigned a good reputation (with probability $\pgc$);
    \item[(2)] interacting with someone who has a good reputation in the eyes of the donor but a bad reputation in the eyes of the observer (with probability $\dtwo$), intending to cooperate, and being assigned a good reputation (with probability $\pbc$);
    \item[(3)] interacting with someone who has a bad reputation in the eyes of the donor but a good reputation in the eyes of the observer (with probability $\dtwo$), intending to defect, and being assigned a good reputation (with probability $\pgd$); or
    \item[(4)] interacting with someone who has a bad reputation in the eyes of both the donor and the observer (with probability $\btwo$, intending to defect, and being assigned a good reputation (with probability $\pbd$).
\end{itemize}
Thus, the probability that a discriminator earns a good reputation is given by
\begin{equation}
    p_\disc = \gtwo \pbc + \dtwo \left(\pbc+\pgd\right) + \btwo \pbd \;.
\end{equation}

\subsection{Agreement and disagreement after gossip}

\textbf{Gossip with a single source.} 
We assume that a single source of gossip is randomly selected after a round of private reputation assessments and that every individual has a probability $\queen$ of consulting the gossip source to (possibly) revise their view of each individual's reputation. The probability that the gossip source views a randomly selected focal individual as good is equivalent to the average reputation $r$ of the population. 

With probability $\queen^2$, then, two individuals randomly selected from the population will have consulted the gossip source (and adopted the source's view) about a focal individual. In this case, the two are guaranteed to agree on the status of the focal individual (view her as good with probability $r$). 
Whereas with probability $1-\queen^2$, at least one of the two will not have consulted the gossip source. In this scenario, the probability that the two individuals agree (or disagree) about the status of the focal individual is identical to the case with fully private information, since we assume that observations are independently (in particular, the gossip source and the two individuals have made independent observations).

In total, then, the probability after the round of gossip that two randomly selected individuals agree a focal individual is good is given by
\begin{equation}
    \gtwo = \left(1-\queen^2\right) \cdot \sum_{\strat\in S} f_\strat r_\strat^2 + \queen^2 \cdot r 
    = \left(1-\queen^2\right) \cdot g_2 + \queen^2 \cdot r \;.
\end{equation}
Similarly, the probability that the two agree that the focal individual is bad is given by
\begin{equation}
    \btwo = \left(1-\queen^2\right) \cdot \sum_{\strat\in S} f_\strat \left(1-r_\strat\right)^2 + \queen^2 \cdot (1-r)
    = \left(1-\queen^2\right) \cdot b_2 + \queen^2 \cdot (1-r) \;. 
\end{equation}
Finally, the probability that the first of the two views the focal individual as good but the second does not is
\begin{equation}
    \dtwo = \left(1-\queen^2\right) \cdot \sum_{\strat\in S} f_\strat r_\strat \left(1-r_\strat\right)
    = \left(1-\queen^2\right) \cdot d_2 \;.
\end{equation}
These quantities satisfy $\gtwo + \btwo + 2\dtwo = 1$, as required.

\textbf{Pairwise gossip with peers (without noise).}
We consider a large, finite population of $N$ individuals engaged in pairwise gossip. In this model, the gossip process for each focal individual $i$ is described by Wright-Fisher process in a population of haploid individuals: at each round of gossip $\gen$, every individual independently and randomly selects a gossip source (equivalent to parentage in the Wright-Fisher model) from the population and adopts the source's view of the focal individual $i$. The two ``alleles'' in the Wright-Fisher model therefore correspond to those individuals who view the focal individual $i$ as good and those who view $i$ as bad. The gossip processes for different focal individuals $i$ are assumed to be independent.

Each gossip process is initialized as follows: at the start of each gossip period ($\gen=0$), we assume that the fraction $r_i$ who view a given focal individual $i$ of type $\strat$ as good is identical to the fraction $r_\strat$ of the population who view type $\strat$ as good in the context of the reputation ODEs (\cref{eq:repODE}). This fraction $r_i$ will be used as the initial ``allele frequency" in the gossip process about individual $i$.

Under this model, the agreement and disagreement terms after $\gen$ Wright-Fisher generations ($N\cdot \gen$ pairwise gossip events) will be
\begin{equation}
    \label{eq:g2b2d2-wf}
    \begin{split}
        \gtwo
        & = \sum_\strat f_\strat \left[ r_\strat^2 + r_\strat \left(1-r_\strat\right) \left(1-\left(1-\dfrac{1}{N}\right)^\gen \right) \right]
        = g_2 + d_2 \left[1-\left(1-\dfrac{1}{N}\right)^\gen \right] 
        \;,
        \\
        \btwo
        & = \sum_\strat f_\strat \left[ \left(1-r_\strat\right)^2 + r_\strat \left(1-r_\strat\right) \left(1-\left(1-\dfrac{1}{N}\right)^\gen \right) \right] 
        = b_2 + d_2 \left[1-\left(1-\dfrac{1}{N}\right)^\gen \right]
        \;,
        \\
        \dtwo
        & = \sum_\strat f_\strat \left[ r_\strat \left(1-r_\strat\right) \left(1-\dfrac{1}{N}\right)^\gen \right] 
        = d_2 \left(1-\dfrac{1}{N}\right)^\gen 
        \;.
    \end{split}
\end{equation}
Assuming $N$ is large but finite, we can use the fact that $(1-1/N)^\gen \approx e^{-\gen/N}$ and let $\sgen\triangleq\gen/N$ to obtain the simplified expressions in \cref{eq:g2b2d2-wf-largeN}.

\subsection{Linear stability analysis} \label{sec:methods-linear}

To determine when gossip can sustain cooperation, we compute the Jacobian of the replicator equations (\cref{eq:replicator}) at the discriminator equilibrium ($f_\disc=1$):
\begin{equation}
    J = \begin{bmatrix}
        (1-\exec) \left( (b r_{\allc}-c)- \left(b-c\right)r_{\disc}) \right) & 0
        \\
        0 & (1-\exec) \left( b r_{\alld} - \left(b-c\right) r_{\disc}) \right)
    \end{bmatrix}
    \Bigg|_{f_{\disc}=1} \;,
\end{equation}
where $r_{\allc}, r_{\alld}, r_{\disc} \in [0,1]$ are evaluated after reputations have reached their equilibrium (i.e., the equilibrium of the ODEs given by \cref{eq:repODE}).
The eigenvalues of $J$, which are simply its diagonal entries here, have no imaginary parts, regardless of the social norm (\SI). Therefore, the discriminator equilibrium is locally stable if and only if the eigenvalues are negative.

We focus on the stability of the discriminator equilibrium in the main text because it is the only equilibrium under Stern Judging and Shunning that supports cooperation (\SI). However, the Simple Standing norm admits a stable mixed equilibrium along the $\allc$--$\disc$ axis, so that cooperation can be sustained as long as an all-$\disc$ population can resist invasion by defectors, i.e., $\lambda_\alld = (1-\exec) \left( b r_{\alld} - \left(b-c\right) r_{\disc}) \right) |_{f_{\disc}=1} < 0$. We visualize this condition in \cref{fig:stabilityALL} in order to facilitate a meaningful comparison across norms.

\subsection{Stochastic simulations} \label{sec:methods-sims}

To verify that our analysis provides a good approximation of a discrete, finite population, we performed a series of Monte Carlo simulations implemented in Julia 1.8.2 \cite{julia_2017}. Each population consists of $N = 100$ individuals, each with a strategy $\strat \in \{\allc, \alld, \disc \} $. Each individual also has a private view of everyone in the population. Generations are partitioned into the following discrete processes, in this order: private assessments, gossip, interactions, and strategy updating.

\textbf{Private assessments.} Each observer $i$ updates their view of each donor $j$ as follows. For each $i, j$ pair, a random recipient $k$ is selected. Each $i$ checks $j$'s most recent action toward $k$ and their own opinion of $k$, then assigns $j$ the corresponding reputational value from a social norm matrix. Then, for each pair $i, j$, a random number is generated; if it is less than $\assess$, the $i$'s view of $j$ is flipped from good to bad or vice versa.

\textbf{Gossip.} The following procedure is iterated $\gen N^3/2$ times. A random triplet $i, j, k$ is chosen. Individual $i$ then adopts $j$'s view of $k$. The $N^3$ comes from rescaling so that one unit of ``time'' corresponds to each individual engaging, on average, in one gossip event; the factor of $1/2$ comes from the fact that heterozygosity decreases twice as quickly in the Moran process as in the Wright-Fisher process used in our analytic treatment.

\textbf{Interactions.} Each donor $i$ interacts with each recipient $j$ according to $i$'s strategy. If $i$ is $\allc$, they cooperate; if $i$ is $\alld$, they defect; and if $i$ is $\disc$, they access their view of recipient $j$, cooperating if that view is good and defecting if it is bad. A random number is selected for each action: if it is less than $\exec$, cooperation is flipped to defection (but not vice versa). Payoffs are updated accordingly: $i$ accrues a benefit $b$ for every co-player who cooperated with $i$ and pays a cost $c$ for every co-player with whom $i$ cooperated.

\textbf{Strategy updating.} A random pair $i, j$ is chosen. Individual $i$ copies $j$'s strategy with probability $1 / (1 + \exp[\omega (\Pi_i - \Pi_j) ])$, where $\Pi_i$ and $\Pi_j$ are their payoffs and $\omega$ is the strength of selection \cite{traulsen_pairwise_2007}; unless otherwise stated, we set $\omega=1$ in our simulations.

We initialize each replicate simulation with a pre-specified number of individuals for each strategy, and with random views and interactions. We then iterate every step of the evolutionary process \emph{except strategy updating} $100$ times, to ensure that reputations and interactions converge to an equilibrium. Finally, we iterate the entire evolutionary process until one strategy has fixed. Example trajectories of strategy frequencies over time are shown in \cref{fig:stabilitySJ}C and D.


\section*{Acknowledgments}

We thank Christian Hilbe for input and discussions about this research.
MK gratefully acknowledges support from James S.~McDonnell Foundation (Postdoctoral Fellowship Award in Understanding Dynamic and Multi-scale Systems, doi:10.37717/2021-3209). JBP and TAK gratefully acknowledge support from the John Templeton Foundation (grant \#62281). 

\addcontentsline{toc}{section}{References}
\setcitestyle{numbers}
\bibliographystyle{apalike}

\setcitestyle{authoryear}

\label{end}

\clearpage
\nolinenumbers
\setcounter{page}{1}
\pagestyle{fancy}
\fancyhf{} 
\renewcommand{\headrulewidth}{0pt}
\fancyfoot[R]{\small Page \thepage \hspace{1pt} of \pageref{LastPage}}

\appendix
\renewcommand{\figurename}{Figure}
\renewcommand{\tablename}{Table}
\renewcommand{\thefigure}{S\arabic{figure}}
\renewcommand{\thetable}{S\arabic{table}}
\renewcommand{\thesection}{S\arabic{section}}
\renewcommand{\theequation}{S\arabic{equation}}
\setcounter{figure}{0}
\setcounter{table}{0}
\setcounter{section}{0}
\setcounter{equation}{0}
\addcontentsline{toc}{section}{Supplementary Information} 
\captionsetup{margin=0.2cm}

{
\vspace*{0.5ex}
\huge \textbf{Supplementary Information}
\vspace{1.5ex}

\LARGE A mechanistic model of gossip, reputations, and cooperation

\large Mari Kawakatsu$^*$, Taylor A.~Kessinger$^*$, Joshua B.~Plotkin

\small $^*$These authors contributed equally
\\
Correspondence to: \href{mailto:marikawa@sas.upenn.edu}{marikawa@sas.upenn.edu} (M.K.), \href{mailto:tkess@sas.upenn.edu}{tkess@sas.upenn.edu} (T.A.K.), \href{mailto:jplotkin@sas.upenn.edu}{jplotkin@sas.upenn.edu}~(J.B.P.)
}

\setcounter{secnumdepth}{3}
\vspace{5ex}

\section{Supplementary Text}

\subsection{Relationship between single-source gossip and empathetic perspective taking}

Here we show the mathematical relationship between the model of gossip with a single source (\cref{fig:schematic}C; \nameref{sec:model}) and the model of empathetic moral evaluation \citeSI{SI_radzvilavicius_evolution_2019}.

The equilibrium of the reputation ODEs, $\frac{dr_\strat}{dt} = p_\strat(t) - r_\strat(t)$ (\cref{eq:repODE}), satisfies $r_\strat = p_\strat$. In particular, the equilibrium reputation of discriminators is
\begin{equation}
    \label{eq:rdisc_SI}
    r_\disc = p_\disc = \gtwo \pgc + \dtwo \left(\pbc + \pgd\right) + \btwo \pbd \;.
\end{equation}
Recall from \cref{eq:g2b2d2-queen} that the rates of agreement and disagreement after a period of gossip are given by 
\begin{equation}
    \begin{split}
        \gtwo & = \left(1-\queen^2\right) \cdot g_2 + \queen^2 \cdot r \;, \\
        \btwo & = \left(1-\queen^2\right) \cdot b_2 + \queen^2 \cdot (1-r) \;, \\
        \dtwo & = \left(1-\queen^2\right) \cdot d_2 \;.
    \end{split}
\end{equation}
Substituting these expressions into \cref{eq:rdisc_SI}, we obtain
\begin{equation}\small
    \begin{split}
        r_\disc & = \bigg[\queen^2 r + \left(1-\queen^2\right) g_2 \bigg] \pgc 
        + \bigg[\left(1-\queen^2\right) d_2 \bigg] \left(\pbc+\pgd\right)
        + \bigg[\queen^2 (1-r) + \left(1-\queen^2\right) b_2 \bigg] \pbd \\
        & = \queen^2 \bigg[ r \pgc + (1-r) \pbd \bigg]
        + \left(1-\queen^2\right) \bigg[ g_2 \pgc + d_2 \left(\pbc + \pgd\right) + b_2 \pbd \bigg] \;.
    \end{split}
\end{equation}
This expression is equivalent in form to Eq.~5 in \citetSI{SI_radzvilavicius_evolution_2019}, but with $\queen^2 = E$, where $E$ is the degree of empathy, i.e., the probability that an observer uses the donor's view of the recipient's reputation when updating the donor's reputation. 
This relationship means that consulting a single gossip source is like a slower form of empathetic perspective-taking: in the former, two individuals will agree with full certainty only if they have both consulted the shared information source, whereas, in the latter, a single individual can guarantee agreement with another by unilaterally adopting her view.

\subsection{Impact of errors in assessment and execution on the critical gossip duration under the Stern Judging norm}
\label{sec:si-errors}

\begin{figure}[h!]
    \centering
    \includegraphics[width=0.6\linewidth]{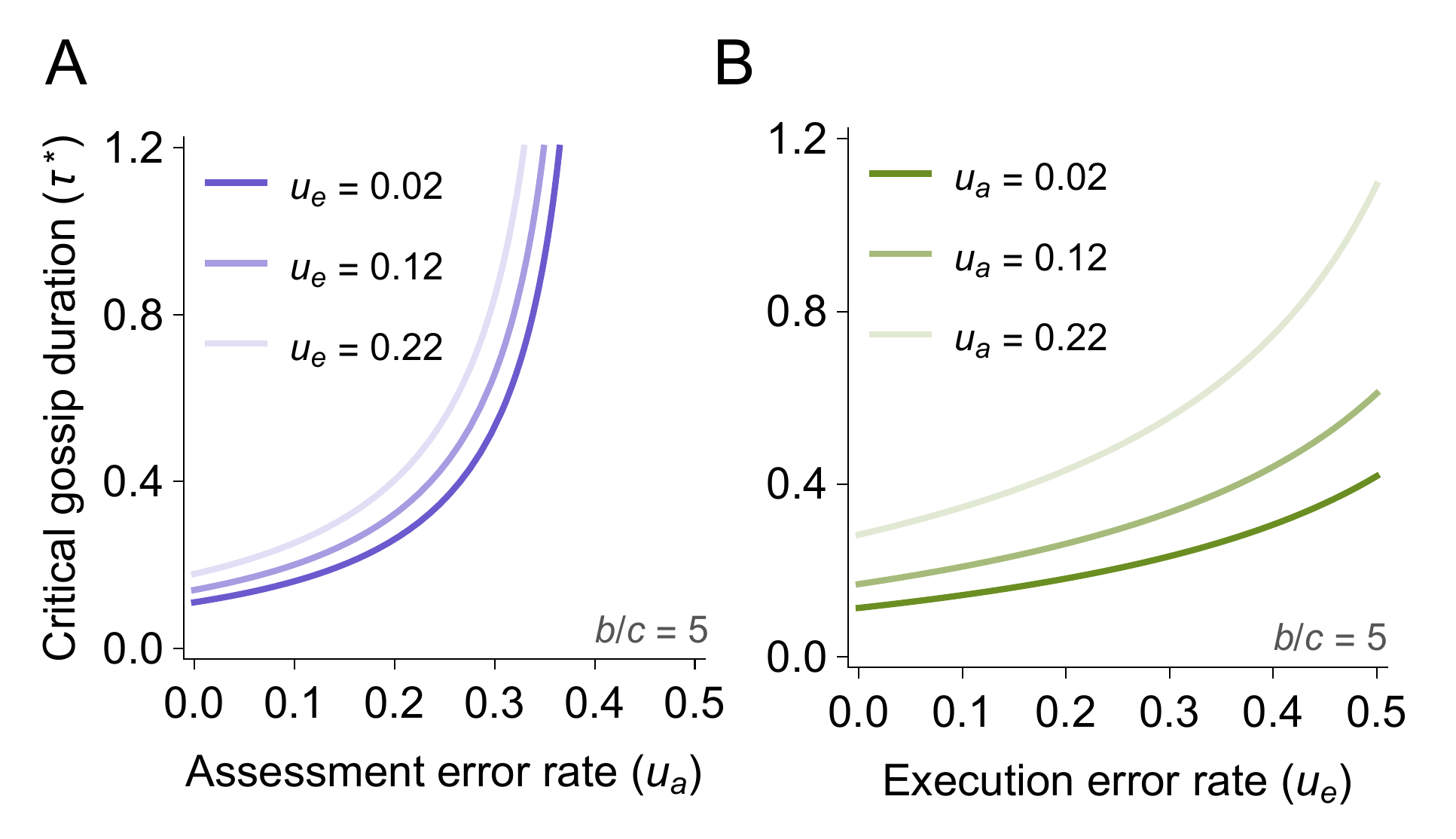}
    \caption{
    \textbf{Impact of assessment and execution errors on the critical gossip duration under the Stern Judging norm.}
    Colors denote assessment error rates $\assess$ (A) or execution error rates $\exec$ (B). For a fixed benefit-to-cost ratio ($b/c=5$), the critical gossip duration $\sgen^*$ increases with either error rate.
    }
    \label{fig:errorsSJ}
\end{figure}

To determine how errors in assessment or execution impact the amount of gossip needed to stabilize cooperation under Stern Judging (condition~(ii) in \maintext), we evaluate the derivatives of the critical gossip duration $\sgen^*$ with respect to the assessment error rate $\assess$ and the execution error rate $\exec$:
\begin{equation}
    \begin{split}
        \dfrac{\partial \sgen^*}{\partial \assess} & 
        = \dfrac{2}{\left(1-2 \assess\right) \left(\frac{\left(b/c\right)}{\left(b/c\right)^*}-1\right)} 
        + \dfrac{\frac{\left(b/c\right)}{\left(b/c\right)^*}}{\left(\frac{1}{\left(b/c\right)^*}-\left(1-\assess\right)\frac{\left(b/c\right)}{\left(b/c\right)^*}\right) \left(2 \left(1-\assess\right)\frac{\left(b/c\right)}{\left(b/c\right)^*}-\frac{1}{\left(b/c\right)^*}\right)} \;,
        \\
        \dfrac{\partial \sgen^*}{\partial \exec} & 
        = \dfrac{1}{(1-\exec) \left(\frac{\left(b/c\right)}{\left(b/c\right)^*}-1\right)} \;,
    \end{split}
\end{equation}
where $\left(b/c\right)^* = \frac{1}{\left(1-2 \assess\right) (1-\exec)}$ as in condition~(i) in \maintext. Both derivatives are positive whenever condition~(i) is satisfied, i.e., $b/c > \left(b/c\right)^*$. Hence, under Stern Judging, the critical gossip duration $\sgen^*$ increases monotonically with $\assess$ and with $\exec$. We confirm this numerically in \cref{fig:errorsSJ}.

\subsection{Equilibrium reputations at the all-DISC equilibrium under any norm} \label{sec:si-eqrep}

To derive conditions for the stability of cooperation, we begin by computing the reputation equilibrium in a population of discriminators.
To do so, we set the right-hand sides of the reputation ODEs (\cref{eq:repODE}) to zero and solve for $r_\allc$, $r_\alld$, and $r_\disc$ at $f_\disc=1$. More explicitly, the reputation equilibrium at the all-$\disc$ equilibrium satisfies
\begin{equation}\small
    \label{eq:si-reps}
    \begin{split}
        r_{\allc} & = r_\disc \pgc + \left(1-r_\disc\right) \pbc \;, \\
        r_{\alld} & = r_\disc \pgd + \left(1-r_\disc\right) \pbd \;, \\
        r_{\disc} & = \left(r_\disc^2+r_\disc\left(1-r_\disc\right)\cdot \left(1-e^{-\sgen}\right) \right) \pgc 
        + \left(r_\disc\left(1-r_\disc\right)\cdot e^{-\sgen} \right) \left(\pbc + \pgd\right) 
        \\
        & \quad\quad
        + \left(\left(1-r_\disc\right)^2+r_\disc\left(1-r_\disc\right)\cdot \left(1-e^{-\sgen}\right) \right) \pbd \;
        \\
        & = \bigg[r_\disc \cdot \pgc + \left(1-r_\disc\right) \cdot \pbd\bigg]
        - e^{-\sgen} \bigg[r_\disc \left(1-r_\disc\right) \cdot \left(\pgc-\pbc-\pgd+\pbd\right) \bigg] \;.
    \end{split}
\end{equation}
These expressions are obtained from \cref{eq:rep_ps} by setting $f_\disc=1$, substituting in the agreement and disagreement rates evaluated at $f_\disc=1$ (\cref{eq:g2b2d2-wf-largeN}), and letting $p_\strat = r_\strat$ (\cref{eq:repODE}).

Solving for $r_\disc$ satisfying $0\leq r_\disc\leq 1$, we obtain the equilibrium reputation of discriminators at the discriminator-only equilibrium:
\begin{equation}\footnotesize
    \begin{split}
    r_\disc = \dfrac{1}{2} \left(
            1 +
            \dfrac{
            e^{\sgen} (1-\pgc+\pbd)}{\pgc-\pbc-\pgd+\pbd}-\sqrt{
                \left(
                    1+\dfrac{e^{\sgen} (1-\pgc+\pbd)}{\pgc-\pbc-\pgd+\pbd}
                \right)^2
                -\dfrac{e^{\sgen}\cdot 4 \pbd }{\pgc-\pbc-\pgd+\pbd}} 
        \right) \;.
    \end{split}
\end{equation}
We can then obtain the equilibrium $r_\allc$ and $r_\alld$ by substituting the expression for $r_\disc$ into the first two equations of \cref{eq:si-reps}.

The explicit expression for the equilibrium value of $r_\disc$ for a norm parametrized by $(\p,\q)$ (\nameref{sec:methods-norms} in \nameref{sec:methods}) can be obtained using \cref{eq:pgcetc}:
\begin{equation}
    \begin{split}
        \pgc & = \left(1-\exec\right)\left(1-\assess\right) + \exec \assess \triangleq \varepsilon \;, \\
        \pgd & = \assess\;, \\
        \pbc & = \p \left(\varepsilon - \assess\right) + \q \left(1 - \varepsilon - \assess\right) + \assess\;,\\
        \pbd & = \q \left(1 - 2 \assess\right) + \assess\;.
    \end{split}
\end{equation}

\textbf{Equilibrium reputations at the all-$\disc$ equilibrium under the Stern Judging norm.} 
The Stern Judging norm is given by $(\p,\q)=(0,1)$, so we have 
$1-\pgc+\pbd = 1+\left(1-2\assess\right)\exec$ and 
$\pgc-\pbc-\pgd+\pbd=2\left(1-2\assess\right)\left(1-\exec\right)$.
The equilibrium reputation of $\disc$ at the all-$\disc$ equilibrium is then
\begin{equation}\small
    \begin{split}
    r_\disc = \dfrac{1}{2} \left(
            1 +
            \dfrac{
            e^{\sgen} (1+\left(1-2\assess\right)\exec)}{2\left(1-2\assess\right)\left(1-\exec\right)}-\sqrt{
                \left(
                    1+\dfrac{e^{\sgen} (1+\left(1-2\assess\right)\exec)}{2\left(1-2\assess\right)\left(1-\exec\right)}
                \right)^2
                -\dfrac{e^{\sgen}\cdot 4\left(1-\assess\right) }{2\left(1-2\assess\right)\left(1-\exec\right)}} 
        \right) \;.
    \end{split}
\end{equation}

\textbf{Equilibrium reputations at the all-$\disc$ equilibrium under the Simple Standing norm.}
The Simple Standing norm is given by $(\p,\q)=(1,1)$, so we have 
$1-\pgc+\pbd = 1+\left(1-2\assess\right)\exec$ and 
$\pgc-\pbc-\pgd+\pbd=\left(1-2\assess\right)\left(1-\exec\right)$.
The equilibrium reputation of $\disc$ at the all-$\disc$ equilibrium is then
\begin{equation}\small
    \begin{split}
    r_\disc = \dfrac{1}{2} \left(
            1 +
            \dfrac{
            e^{\sgen} (1+\left(1-2\assess\right)\exec)}{\left(1-2\assess\right)\left(1-\exec\right)}-\sqrt{
                \left(
                    1+\dfrac{e^{\sgen} (1+\left(1-2\assess\right)\exec)}{\left(1-2\assess\right)\left(1-\exec\right)}
                \right)^2
                -\dfrac{e^{\sgen}\cdot 4\left(1-\assess\right) }{\left(1-2\assess\right)\left(1-\exec\right)}} 
        \right) \;.
    \end{split}
\end{equation}

\textbf{Equilibrium reputations at the all-$\disc$ equilibrium under the Shunning norm.}
The Shunning norm is given by $(\p,\q)=(0,0)$, so we have 
$1-\pgc+\pbd = \left(1-2\assess\right)\exec+2\assess$ and 
$\pgc-\pbc-\pgd+\pbd=\left(1-2\assess\right)\left(1-\exec\right)$.
The equilibrium reputation of $\disc$ at the all-$\disc$ equilibrium is then
\begin{equation}\small
    \begin{split}
    r_\disc = \dfrac{1}{2} \left(
            1 +
            \dfrac{
            e^{\sgen} (\left(1-2\assess\right)\exec+2\assess)}{\left(1-2\assess\right)\left(1-\exec\right)}-\sqrt{
                \left(
                    1+\dfrac{e^{\sgen} (\left(1-2\assess\right)\exec+2\assess)}{\left(1-2\assess\right)\left(1-\exec\right)}
                \right)^2
                -\dfrac{e^{\sgen}\cdot 4\assess }{\left(1-2\assess\right)\left(1-\exec\right)}} 
        \right) \;.
    \end{split}
\end{equation}

\subsection{Stability of the all-DISC equilibrium against ALLD under any norm}

We focus on the stability of the $\disc$ equilibrium in the main text because it is the only equilibrium under the Stern Judging norm that stably supports cooperation (this is also the case under the Shunning norm). However, the Simple Standing norm admits a stable mixed equilibrium along the $\allc$--$\disc$ axis, such that cooperation can be sustained even when an all-discriminator population can be invaded by defectors. To facilitate meaningful comparisons across norms, we analyze the stability of the all-discriminator equilibrium against only $\alld$ (vs against both $\alld$ and $\allc$ as in the analysis under Stern Judging reported in \maintext).

The replicator dynamic ODE for the competition between $\alld$ and $\disc$ is given by
\begin{equation}
    \dfrac{df_\disc}{d\eta} = f_\disc \left(1-f_\disc\right) \left(\pi_\disc-\pi_\alld\right) \;.
\end{equation}
Here, $r_{\alld}$, $r_{\disc} \in [0,1]$ are evaluated at the reputation equilibrium, as before (when $f_\disc=1$, the equilibrium reputations in the two-strategy case are identical to the three-strategy case analyzed in Section~\ref{sec:si-eqrep} above). 

The Jacobian of this ODE at the all-discriminator equilibrium ($f_\disc=1$) is given by
\begin{equation}
    \label{eq:si-jacobian}
    J = (1-\exec) \left( b r_{\alld} - \left(b-c\right) r_{\disc}) \right) \big|_{f_{\disc}=1} \;.
\end{equation}

The all-discriminator equilibrium is locally stable if and only if $J<0$.
For a general norm parametrized by $\left(\p,\q\right)$ (see \nameref{sec:methods-norms} in \nameref{sec:methods}), this condition simplifies to
\bgroup\scriptsize
\begin{align}
    & \textrm{(i$'$)  }\quad \dfrac{b}{c} > \left(\dfrac{b}{c}\right)^* 
    = \dfrac{1}{\left(1-2 \assess\right) (1-\exec)}
    \qquad\textrm{and}\qquad
    \\
    & \textrm{(ii$'$)  }\quad 
    \begin{cases} \sgen
    > \sgen^*_\alld
    = \log \left[
    (1-\p+\q) 
    \left(1- \dfrac{\assess + \q (1 - 2 \assess)}{1 + \q (1 - 2 \assess)} \cdot
    \dfrac{\left(\dfrac{b}{c}\right)}{\left(\dfrac{b}{c}\right)-\dfrac{1}{1+q (1-2 \assess)}}\right)
    \left(\dfrac{\dfrac{b}{c}}{\dfrac{b}{c}-\left(\dfrac{b}{c}\right)^*}\right)
    \right]
    & \text{ if }\ \sgen^*_\alld \geq 0
    \;,
    \\
    \sgen
    \geq 0 
    & \text{ if }\ \sgen^*_\alld < 0 \;.
    \end{cases}
\end{align}
\egroup

Note that $\sgen^*_\alld$ is undefined for the Scoring norm ($(\p,\q)=(1,0)$). This is consistent with the fact that, under a first-order norm, gossip will not impact the stability of $\disc$ because equilibrium reputations do not depend on the level of agreement about social reputations. For any norm other than Scoring (i.e., $(\p,\q)\in[0,1]^2\setminus(1,0)$), $\sgen^*_\alld$ is a decreasing function of $b/c$ for all $(b/c)>(b/c)^*$ (i.e., when (i$'$) is satisfied), meaning that less gossip is required to stabilize cooperation when the benefit-to-cost ratio is larger.

\begin{figure}[h!]
    \centering
    \includegraphics[width=0.97\linewidth]{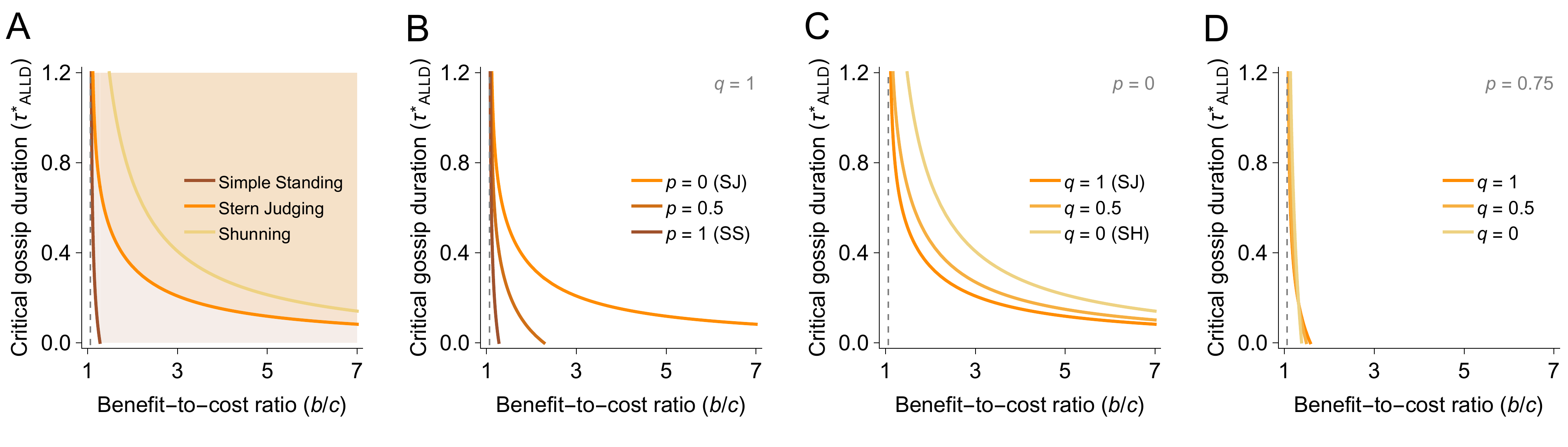}
    \caption{
        \textbf{Impact of social norm on the critical gossip duration for $\disc$ to resist $\alld$.}
        Panels show the critical gossip duration $\sgen_{\alld}^*$ for a population of discriminators ($\disc$) to resist invasion by defectors ($\alld$) as a function of the benefit-to-cost ratio. Colors denote social norms, parameterized by the probability $\p$ ($\q$) that cooperating with (defecting against) a bad recipient yields a good reputation.
        \textbf{A}: For a given benefit-to-cost ratio $b/c$, the critical threshold $(\sgen)_{\alld}^*$ is the smallest for Simple Standing (SS; $(\p,\q)=(1,1)$), intermediate for Stern Judging (SJ; $(\p,\q)=(0,1)$), and the largest for Shunning (SH; $(\p,\q)=(0,0)$).
        \textbf{B}: The critical gossip duration decreases with increasing $\p$, which makes a norm more `lenient' (i.e., incentivizes cooperating with `bad' individuals). Parameter $\q=1$ is fixed.
        \textbf{C, D}: Depending on parameter values, the critical gossip duration can increase or decrease with increasing $\q$, which makes a norm more `strict' (i.e., incentivizes punishing `bad' individuals). Parameter $\p$ is fixed: $p=0$ (C) and $p=0.75$ (D).
        Other parameters: $\assess=\exec=0.02$.
    }
    \label{fig:stabilityALL}
\end{figure}

\textbf{The effect of the social norm on the critical gossip duration.} We evaluate conditions (i$'$) and (ii$'$) for the three second-order norms of interest: Stern Judging ($(\p,\q)=(0,1)$), Simple Standing ($(\p,\q)=(1,1)$), and Shunning ($(\p,\q)=(0,0)$) (\cref{fig:stabilityALL}A). Consistent with our intuition, the duration of gossip $\sgen^*_\alld$ needed to stabilize $\disc$ against $\alld$ is the highest for the Shunning norm, the lowest for the Simple Standing norm, and intermediate for the Stern Judging norm.

The conditions above also allow us to study the impact of a general social norm $(\p,\q)$ on the critical gossip duration. We find analytically that the critical gossip duration $\sgen^*_\alld$ is decreasing in $\p$ (i.e., $\partial\sgen^*_\alld/\partial\p < 0$ for any $b>c>0$ and $0<\assess,\exec<1/2$; see a numerical example in \cref{fig:stabilityALL}B). This is consistent with the intuition that increasing the parameter $\p$ makes the norm more `lenient', incentivizes cooperating with bad individuals, and therefore reduces the amount of gossip needed to stabilize cooperation. 

In contrast, we find that $\sgen^*_\alld$ is increasing or decreasing in $\q$ depending on parameter conditions: if condition (i$'$) is satisfied ($(b/c)>(b/c)^*$), then 
\bgroup\small
\begin{equation}
    \dfrac{\partial\sgen^*_\alld}{\partial\q} > 0
    \ \Longleftrightarrow\ 
    \assess\geq \dfrac{1-\exec}{2 (2-\exec)}\ \textrm{or}\  \assess>\dfrac{1}{2 b}\ \textrm{or}\  \left(\assess=\dfrac{1}{2 b}\ \textrm{and}\  p>0\right)\ \textrm{or}\  \left(\assess<\dfrac{1}{2 b}\ \textrm{and}\  p>\dfrac{1-2 b \assess}{b (1-2 \assess)}\right)\;.
\end{equation}
\egroup
Numerical examples in \cref{fig:stabilityALL}C and D are consistent with these analytical results. Increasing the parameter $\q$ generally makes the norm more `strict' and incentivizes defecting against bad individuals. This can in turn promote cooperation and thus lower the critical gossip duration, at least when assessments are relatively accurate (low $\assess$) and cooperating with bad individuals is disincentivized (low $\p$) (\cref{fig:stabilityALL}C). When $\p$ is high(er), however, this effect is reversed for some combinations of the benefit-to-cost ratio $b/c$ and the assessment error rate $\assess$ (\cref{fig:stabilityALL}D).

\textbf{The effect of assessment and execution errors on the critical gossip duration.}
To determine how errors in assessment or execution impact the amount of gossip needed to stabilize a population of $\disc$ against $\alld$, we evaluate the derivatives of the critical gossip duration $\sgen^*_\alld$ with respect to the assessment error rate $\assess$ and the execution error rate $\exec$.

\begin{figure}[!h]
    \centering
    \includegraphics[width=0.8\linewidth]{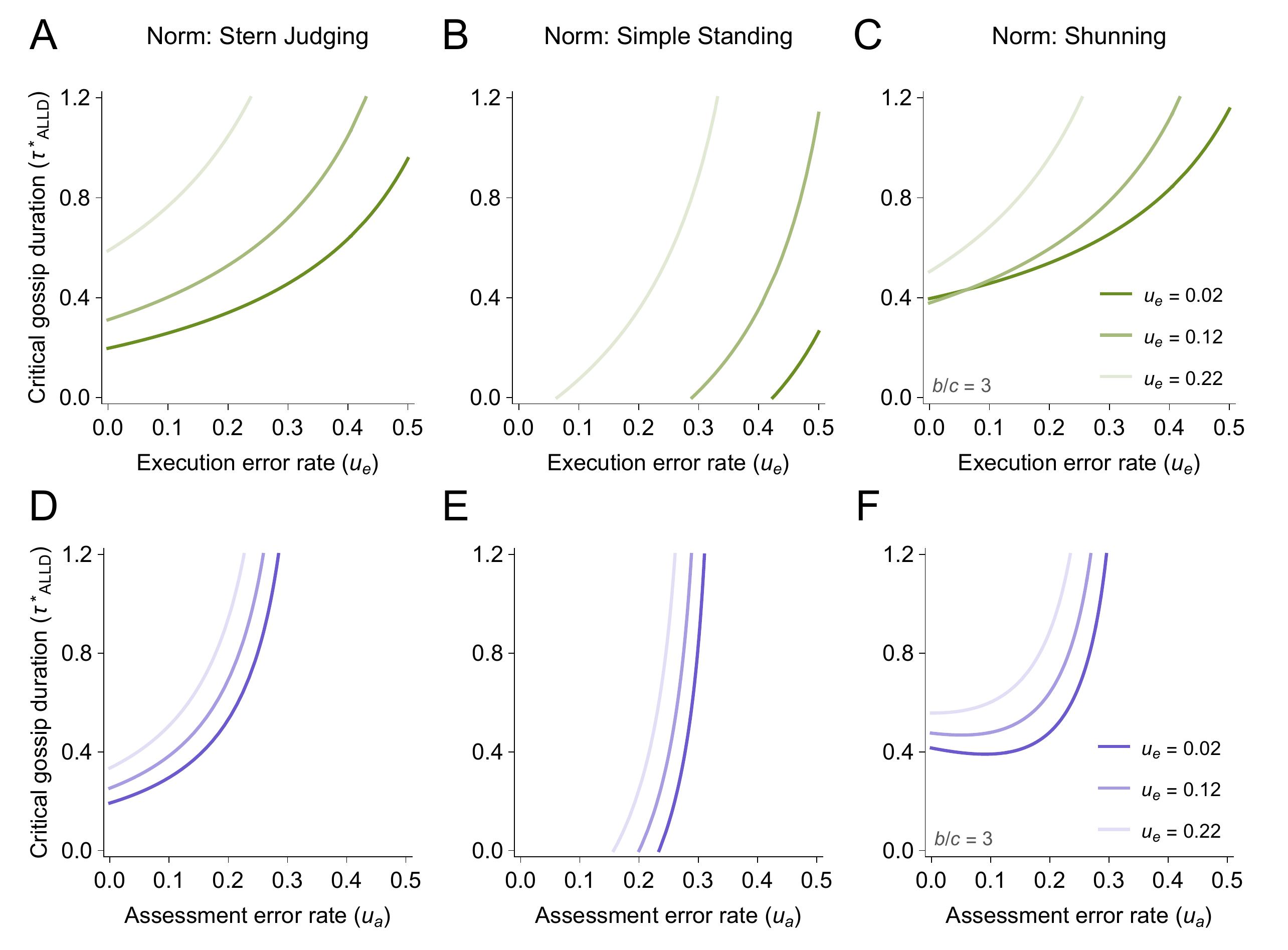}
    \caption{
        \textbf{Impact of errors on the critical gossip duration for $\disc$ to resist $\alld$.}
        Colors denote execution error rates $\exec$ (A--C) or assessment error rates $\assess$ (D--F). 
        \textbf{A--C}: For a fixed benefit-to-cost ratio ($b/c=3$), the critical gossip duration $\sgen^*_\alld$ increases with increasing $\exec$.
        \textbf{D--F}: For a fixed benefit-to-cost ratio ($b/c=3$), the critical gossip duration $\sgen^*_\alld$ increases with increasing $\assess$ under Stern Judging and Simple Standing, but $\sgen^*_\alld$ is non-monotonic in $\assess$ under Shunning.
    }
    \label{fig:errorsALL}
\end{figure}

The critical gossip duration $\sgen^*_\alld$ is increasing in $\exec$ (see numerical examples in \cref{fig:errorsALL}A--C): if condition (i$'$) is satisfied ($(b/c)>(b/c)^*$), then we have
\begin{equation}
    \dfrac{\partial\sgen^*_\alld}{\partial\exec} 
    = \dfrac{\left(b/c\right)^*}{(1-\exec) \left(\left(b/c\right)-\left(b/c\right)^*\right)}
    > 0
\end{equation}
for any $b>c>0$ and $0<\assess,\exec<1/2$.

In contrast, $\sgen^*_\alld$ is increasing or decreasing in $\assess$ depending on the social norm and parameter values. Under Stern Judging and Simple Standing, $\sgen^*_\alld$ is increasing in $\assess$ (i.e., $\partial\sgen^*_\alld/\partial\assess > 0$ for any $b>c>0$ and $0<\assess,\exec<1/2$ (numerical examples in \cref{fig:errorsALL}D and E). However, under Shunning, $\sgen^*_\alld$ can be monotonic in $\assess$ (numerical example in \cref{fig:errorsALL}F): we have
\begin{equation}
    \dfrac{\partial\sgen^*_\alld}{\partial\assess} > 0
    \quad\Longleftrightarrow\quad
    \dfrac{b}{c} < \dfrac{4}{3-4 \assess-\sqrt{1+ 8 \exec+ 8 \assess (1-2 \assess-4 (1-\assess) \exec)}} 
\end{equation}
assuming condition (i$'$) is satisfied ($(b/c)>(b/c)^*$).
For $\assess=\exec=0.02$, the condition on the right-hand side evaluates to $(1.06293=(b/c)^*<) b/c<2.248$.

\newpage
\subsection{Agreement and disagreement after gossip (with bias)} 

Next, we derive the expressions for the agreement and disagreement terms, $\gtwo$, $\btwo$, and $\dtwo$, after biased gossip (\cref{eq:g2b2d2-wf-bias}).

\textbf{Gossip process for a focal individual.} 
We consider a population of $N$ individuals engaged in gossip. Suppose that, at time $\gen$, there are $\ell$ individuals who believe a focal individual $i$ is good and $N-\ell$ who believe $i$ is bad. We assume that, with probability $\GtoB$ ($\BtoG$), an individual who considered $i$ as good (bad) ``mutates'' to the opposite opinion between time $\gen$ and $\gen+1$. Thus, the dynamics of biased gossip for a focal individual follow a Wright-Fisher process in a haploid population with two alleles, which keeps track of how many individuals view the focal individual as good (allele one) or bad (allele two) over discrete generations (rounds) of gossip.

Let $R_{i,\gen}\in\{0,1/N,\dots,\left(N-1\right)/N,1\}$ be a random variable that tracks the frequency of allele one at time $\gen$ (after $\gen$ rounds of gossip).
The probability that there are $m$ individuals who believe $i$ is good at time $\gen+1$ is
\begin{equation}
    \begin{split}
        p_{m\ell} = 
        \mathbb{P}\left(R_{i,\gen+1} = \dfrac{m}{N}\ \bigg|\ R_{i,\gen} = \dfrac{\ell}{N} \right)
        & = \binom{N}{m} \left(g\left(\dfrac{\ell}{N}\right)\right)^m \left(1-g\left(\dfrac{\ell}{N}\right)\right)^{N-m}
    \end{split}
\end{equation}
for $0\leq m \leq N$, where the function
\begin{equation}
    g(r_{i,\gen}) 
    = r_{i,\gen} \left(1-\GtoB\right) + \left(1-r_{i,\gen}\right) \BtoG
    = \left(1-\GtoB-\BtoG\right)r_{i,\gen} + \BtoG 
\end{equation}
gives the proportion of gossip transmitted between $\gen$ and $\gen+1$ that is positive (i.e., views $i$ as good), provided that a fraction $r_{i,\gen}$ view $i$ as good at time $\gen$. In the absence of mutation ($\GtoB=\BtoG=0$), we recover the model of gossip as pure drift ($g(r_{i,\gen})=r_{i,\gen}$). 

The mean and variance of the distribution of $R_{i,\gen}$ are given by \citeSI[see][]{SI_tataru_inference_2015,SI_tataru_statistical_2017}
\begin{equation}\small
    \begin{split}
        \E\left[R_{i,\gen} \ \big|\ R_{i,0} = r_{i,0} \right] 
        & = \left(r_{i,0}-\dfrac{\BtoG}{\GtoB+\BtoG}\right) \left(1-\GtoB-\BtoG\right)^\gen + \dfrac{\BtoG}{\GtoB+\BtoG} 
        \;,
        \\
        \Var \left(R_{i,\gen} \big| R_{i,0} = r_{i,0} \right)
        & = \dfrac{\BtoG}{\GtoB+\BtoG} \left(1-\dfrac{\BtoG}{\GtoB+\BtoG}\right) 
        \left[ \dfrac{1 - \left(1-\frac{1}{N}\right)^\gen \left(1-\left(\GtoB+\BtoG\right)\right)^{2\gen} }{ N - \left(N-1\right) \left(1-\left(\GtoB+\BtoG\right)\right)^{2} }\right]
        \\ & \quad + \left(1-2\cdot \dfrac{\BtoG}{\GtoB+\BtoG}\right) \left(r_{i,0}-\dfrac{\BtoG}{\GtoB+\BtoG}\right) \left(1-\left(\GtoB+\BtoG\right)\right)^\gen 
        \left[ \dfrac{1 - \left(1-\frac{1}{N}\right)^\gen \left(1-\left(\GtoB+\BtoG\right)\right)^{\gen} }{ N - \left(N-1\right) \left(1-\left(\GtoB+\BtoG\right)\right) }\right]
        \\ & \quad - \left(r_{i,0}-\dfrac{\BtoG}{\GtoB+\BtoG}\right)^2 \left(1-\left(\GtoB+\BtoG\right)\right)^{2\gen} \left[ 1 - \left(1-\dfrac{1}{N}\right)^\gen \right]
        \;.
    \end{split}
\end{equation}
Assuming (1) $N$ is large and (2) $\GtoB$ and $\BtoG$ are small, we can approximate these quantities as 
\begin{equation}\small
    \begin{split}
        \E\left[R_{i,\sgen} \ \big|\ R_{i,0} = r_{i,0} \right] 
        & = \left(r_{i,0}-\dfrac{\BTOG}{\GTOB+\BTOG}\right) e^{-\left(\GTOB+\BTOG\right)\sgen} + \dfrac{\BTOG}{\GTOB+\BTOG} 
        \;,
        \\
        \Var \left(R_{i,\sgen} \big| R_{i,0} = r_{i,0} \right)
        & = \dfrac{\BTOG}{\GTOB+\BTOG} \left(1-\dfrac{\BTOG}{\GTOB+\BTOG}\right) 
        \cdot
        \dfrac{1}{1 + 2\left(\GTOB+\BTOG\right) } \left(1 - e^{-\left(2\left(\GTOB+\BTOG\right)+1\right)\sgen}\right)
        \\ & \qquad + \left(1-\dfrac{2\BTOG}{\GTOB+\BTOG}\right) \left(r_{i,0}-\dfrac{\BTOG}{\GTOB+\BTOG}\right) \dfrac{1}{ 1 + \left(\GTOB+\BTOG\right)}
        \cdot 
        e^{-\left(\GTOB+\BTOG\right)\sgen} \left(1 - e^{-\left(\left(\GTOB+\BTOG\right)+1\right)\sgen}\right)
        \\ & \qquad - \left(r_{i,0}-\dfrac{\BTOG}{\GTOB+\BTOG}\right)^2 e^{-2\left(\GTOB+\BTOG\right)\sgen} \left( 1 - e^{-\sgen} \right)
        \;,
    \end{split}
\end{equation}
where $\GTOB=N\GtoB$ and $\BTOG=N\BtoG$ are the scaled mutation rates and $\sgen=\gen/N$ is the scaled gossip duration. We refer the reader to \citetSI{SI_tataru_inference_2015,SI_tataru_statistical_2017} for the derivations.

\textbf{Population-level agreement and disagreement.}
We derive the agreement and disagreement terms, $\gtwo$, $\btwo$, and $\dtwo$, by first computing the following quantities for a focal individual $i$:
\begin{equation}
    \begin{split}
        \E\left[R_{i,\sgen}^2 \big| R_{i,0}=r_{i,0} \right] 
        & = \Var \left(R_{i,\sgen} \big| R_{i,0}=r_{i,0} \right) + \E\left[R_{i,\sgen} \big| R_{i,0}=r_{i,0} \right]^2 \;,
        \\
        \E\left[(1-R_{i,\sgen})^2 \big| R_{i,0}=r_{i,0} \right]
        & = 1 - 2 \E\left[R_{i,\sgen} \big| R_{i,0}=r_{i,0} \right] + \E\left[R_{i,\sgen}^2 \big| R_{i,0}=r_{i,0} \right]  \;,
        \\
        \E\left[R_{i,\sgen}(1-R_{i,\sgen}) \big| R_{i,0}=r_{i,0} \right]
        & = \E\left[R_{i,\sgen} \big| R_{i,0}=r_{i,0} \right] - \E\left[R_{i,\sgen}^2 \big| R_{i,0}=r_{i,0} \right]  \;.
    \end{split}
\end{equation}

As discussed in the main text (\nameref{sec:result3}), the gossip process for a focal individual $i$ is initialized as follows: at the start of each gossip period ($\gen=0$), we assume that the fraction $r_i$ of those engaged in gossip who view a given focal individual $i$ of type $\strat$ as good is identical to the fraction $r_\strat$ of the population who view type $\strat$ as good in the context of the reputation ODEs (\cref{eq:repODE}). In other words, the initial allele one frequency in the gossip process about individual $i$, $r_{i,0}$, is $r_\strat$. Therefore, the quantities $\gtwo$, $\btwo$, and $\dtwo$ can be computed as
\begin{equation}
    \begin{split} 
        \gtwo
        & = \sum_\strat f_\strat \cdot \E\left[R_{i,\sgen}^2 \big| R_{i,0}=r_\strat \right] 
        \;,
        \\
        \btwo 
        & = \sum_\strat f_\strat \cdot \E\left[(1-R_{i,\sgen})^2 \big| R_{i,0}=r_\strat \right]
        \;,
        \\
        \dtwo 
        & = \sum_\strat f_\strat \cdot \E\left[R_{i,\sgen}(1-R_{i,\sgen}) \big| R_{i,0}=r_\strat \right]
        \;,
    \end{split}
\end{equation}
as reported in \cref{eq:g2b2d2-wf-bias}.


\addcontentsline{toc}{section}{References}
\setcitestyle{numbers}

\clearpage
\section{Supplementary Figures}

\vspace{0.75in}

\begin{figure}[h!]
    \centering
    \includegraphics[width=0.4\textwidth,trim={0cm 0 -1.5cm 0cm},clip]{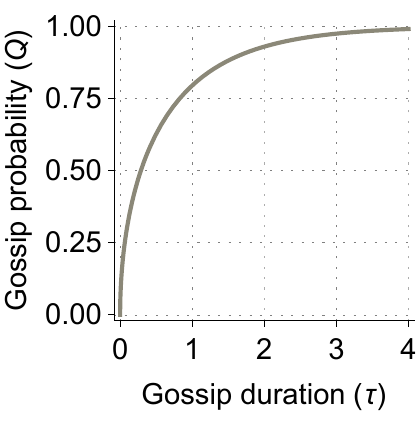}
    \caption{
    \textbf{Relationship between gossip with a single source versus peer-to-peer gossip.} 
    These two distinct gossip processes have the same effects on the level of agreement and equilibrium reputations in the population under a suitable transformation of parameters. We plot the transformation $\sgen=-\log\left(1-\queen^2\right)$ between the duration of gossip $\sgen$ in the peer-to-peer process and the probability $\queen$ of consulting the single gossip source (\cref{eq:mapping}).
    }
    \label{fig:mapping}
\end{figure}

\vspace{0.75in}

\begin{figure}[h!]
    \centering
    \includegraphics[width=0.97\linewidth]{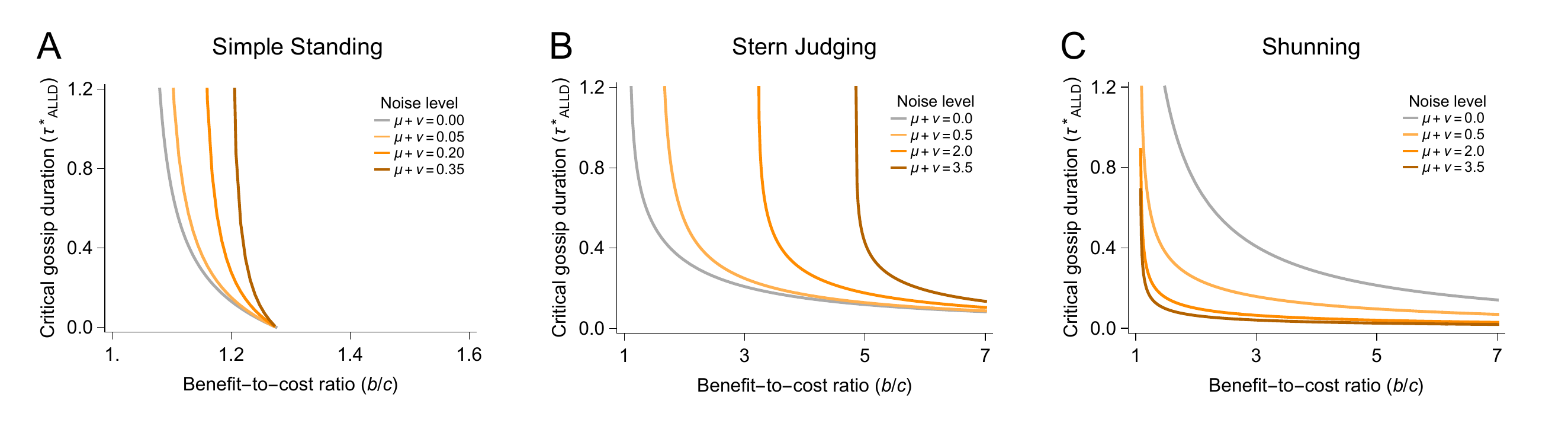}
    \caption{
        \textbf{Effects of noisy gossip on cooperation.}
        Panels show the critical gossip duration $\sgen^*_\alld$ for a population of discriminators ($\disc$) to resist invasion by defectors ($\alld$) as a function of the benefit-to-cost ratio, under the Simple Standing (A), Stern Judging (B), and Shunning (C) norms.
        Colors denote different amounts of unbiased noise in gossip ($\GTOB+\BTOG$). Each gray line indicates the critical gossip duration for noiseless transmission ($\GTOB\!=\!\BTOG\!=\!0$) under the corresponding norm.
        The critical threshold $\sgen_{\alld}^*$ increases with noise under Simple Standing (A) and Stern Judging (B; see also \cref{fig:stabilitySJ}A), but the trend reverses under Shunning (C). Under the Shunning norm, reputations (before gossip) are overwhelmingly negative, and this negativity tends to be self-reinforcing because donors who cooperate with bad individuals themselves gain bad reputations; noisy gossip helps break this cycle by stochastically introducing positive gossip and, consequently, makes it easier to sustain cooperation. 
        Other parameters: $\assess=\exec=0.02$.
        Note that panel B is identical to \cref{fig:noiseSJ}A (i.e., $\sgen^*=\sgen^*_\alld$ under the Stern Judging norm) and is shown again here to facilitate comparison across norms.
    }
    \label{fig:noiseALL}
\end{figure}

\begin{figure}[h!]
    \centering
    \includegraphics[width=0.85\linewidth]{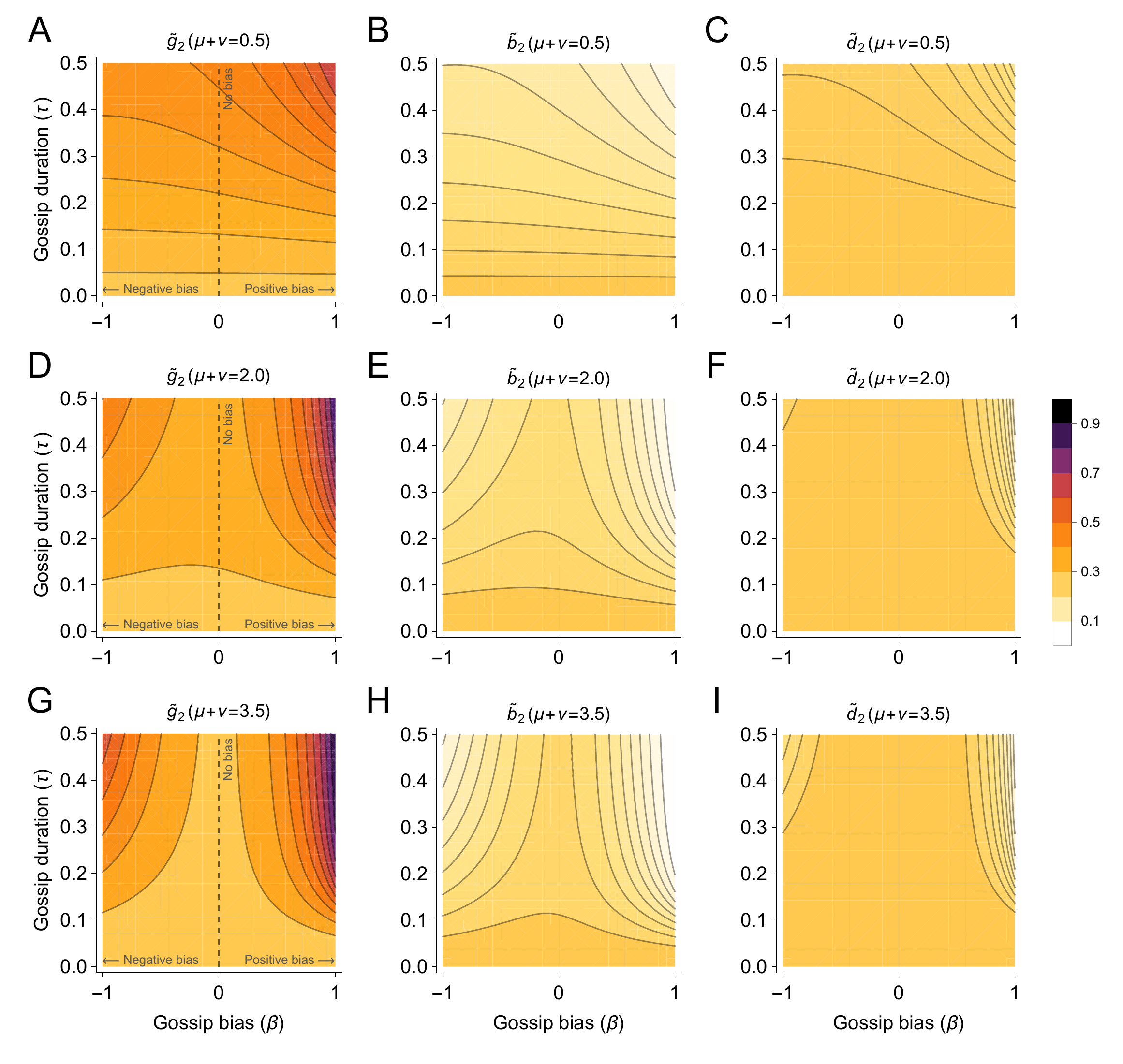}
    \caption{
    \textbf{Agreement and disagreement at the discriminator-only equilibrium under the Stern Judging norm.}
    We plot the agreement ($\gtwo$, $\btwo$) and disagreement ($\dtwo$) terms for $f_\disc=1$ at the reputation equilibrium as a function of gossip duration $\sgen$ and gossip bias $\beta$ in different mutation regimes ($\GTOB+\BTOG=0.5$ in A--C, $2.0$ in D--F, $3.5$ in G--I).
    The terms were computed as $\gtwo=\sum_\strat f_\strat r_\strat^2=r_\disc^2$, $\btwo=\sum_\strat f_\strat \left(1-r_\strat\right)^2=\left(1-r_\disc\right)^2$, and $\dtwo=\sum_\strat f_\strat r_\strat \left(1-r_\strat\right)=r_\disc \left(1-r_\disc\right)$.
    Darker colors indicate greater levels of agreement (for $\gtwo$ and $\btwo$; A,~D,~G and B,~E,~H) or disagreement (for $\btwo$; C,~F,~I).
    Other parameters: $\assess=\exec=0.02$.
    }
    \label{fig:agreementSJ}
\end{figure}

\begin{figure}[h!]
    \centering
    \includegraphics[width=0.85\linewidth]{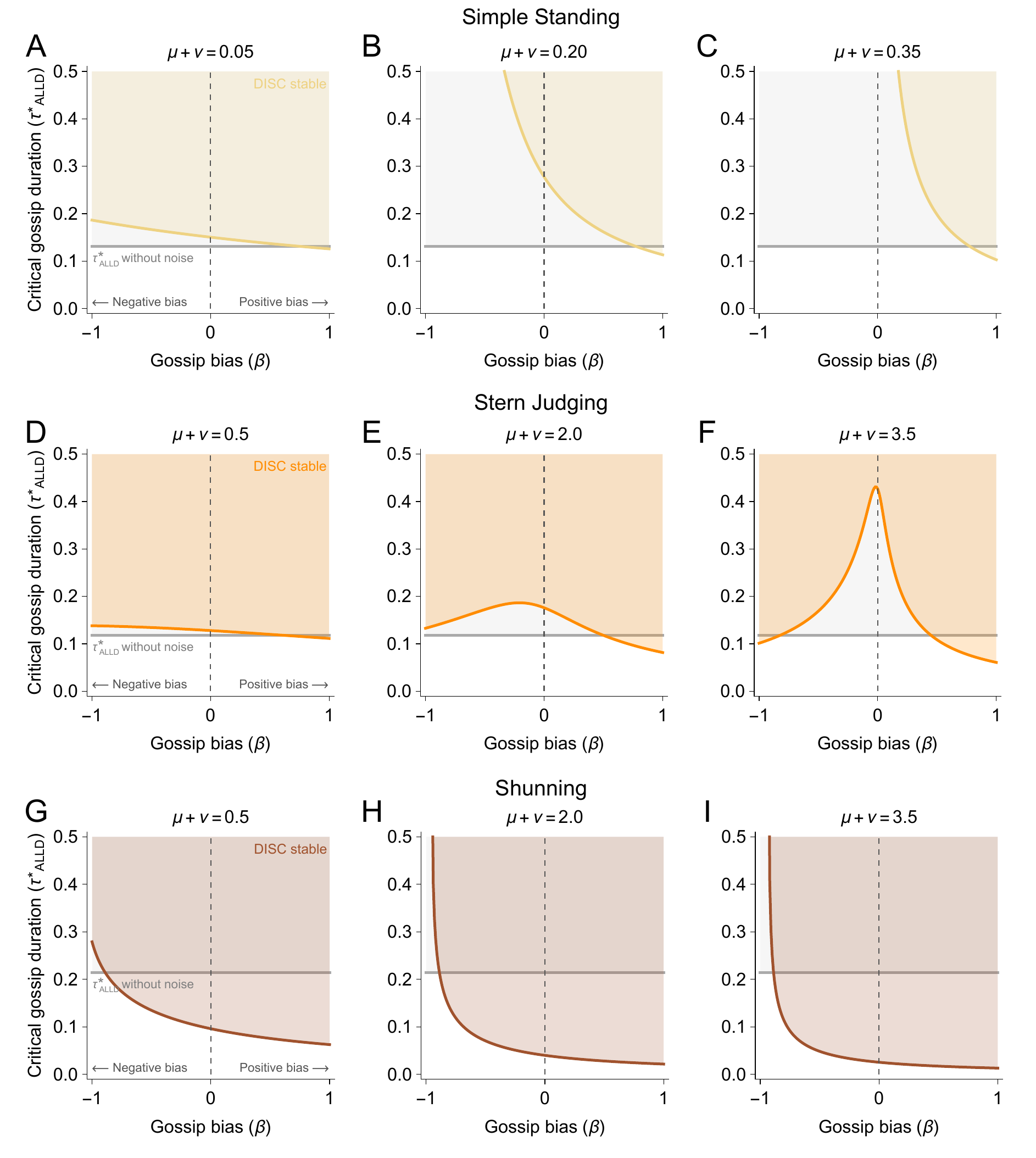}
    \caption{
        \textbf{Effects of biased gossip on cooperation.}
        Panels show the critical gossip duration $\sgen^*_\alld$ for a population of discriminators ($\disc$) to resist invasion by defectors ($\alld$) as a function of gossip bias ($\beta$), under the Simple Standing (A--C), Stern Judging (D--F), and Shunning (G--I) norms. 
        Columns denote different regimes of noise as indicated.
        Solid gray lines (identical across panels within each row) indicate the baseline critical gossip duration $\sgen^*_\alld$ in the absence of transmission noise ($\GTOB=\BTOG=0$).
        Parameters: $b/c=5$, $\assess=\exec=0.02$.
        Note that panels D--F are identical to \cref{fig:biasSJ}A--C (i.e., $\sgen^*=\sgen^*_\alld$ under the Stern Judging norm) and are shown again here to facilitate comparison across norms.
    }
    \label{fig:biasALL}
\end{figure}

\end{document}